\documentclass[pre,twocolumn,superscriptaddress,floatfix]{revtex4-1}
\usepackage[latin1]{inputenc}
\usepackage{graphicx}
\usepackage{amsmath}
\usepackage{amssymb}
\usepackage{color}
\usepackage[dvips]{epsfig}
\usepackage{epstopdf}
\usepackage{times,mathptmx}
\usepackage{epstopdf}
\usepackage{natbib}

\DeclareMathOperator{\arsinh}{arsinh}

\begin{document}
\title{Enhanced transport of ions  by tuning surface properties of the nanochannel}
%\title{Amplification of ion transport in nanochannels by tuning surface properties  }
\author{Olga I. Vinogradova}
\email[Corresponding author: ]{oivinograd@yahoo.com}
\affiliation{Frumkin Institute of Physical Chemistry and
Electrochemistry, Russian Academy of Science, 31 Leninsky Prospect,
119071 Moscow, Russia}
\author{Elena F. Silkina}
\affiliation{Frumkin Institute of Physical Chemistry and
Electrochemistry, Russian Academy of Science, 31 Leninsky Prospect,
119071 Moscow, Russia}
\author{Evgeny S. Asmolov}
\affiliation{Frumkin Institute of Physical Chemistry and
Electrochemistry, Russian Academy of Science, 31 Leninsky Prospect,
119071 Moscow, Russia}

\date{\today }

\begin{abstract}
Motivated by recent observations of anomalously large deviations of the conductivity currents in confined systems from the bulk behavior, we revisit the theory of ion transport in parallel-plate channels and also discuss how the wettability of a solid and the mobility of adsorbed surface charges impact the transport of ions. It is shown that depending on the ratio of the electrostatic disjoining pressure to the excess osmotic pressure at the walls two different regimes occur. In the thick channel regime this ratio is small and the channel effectively behaves as thick, even when the diffuse layers strongly overlap. The latter is possible for highly charged channels only. In the thin channel regime the disjoining pressure is comparable to the excess osmotic pressure at the wall, which implies relatively weakly charged walls. We derive simple expressions for the mean conductivity of the channel in these two regimes, highlighting the role of electrostatic and electro-hydrodynamic boundary conditions. Our theory provides a simple explanation of the high conductivity observed experimentally in hydrophilic channels,  and allows one to obtain rigorous bounds on its attainable value and scaling with salt concentration. Our results also show that further dramatic amplification of conductivity is possible if hydrophobic slip is involved, but only in the thick channel regime provided the walls are sufficiently highly charged and the most of adsorbed charges are immobile. However, for weakly charged surfaces the massive conductivity amplification due to hydrodynamic slip is impossible in both regimes. Interestingly, in this case the moderate slip-driven contribution to conductivity can monotonously decrease with the fraction of immobile adsorbed charges.
These results  provide a framework for tuning the conductivity of nanochannels by adjusting their surface properties and bulk electrolyte concentrations.
\end{abstract}

\maketitle

\section{Introduction}\label{sec:intro}

Surface conductivity is a name given to an extra conductivity within an electrostatic diffuse layer (EDL),  i.e. the region, where the surface charge is balanced by a cloud of counter-ions,  compared to its value in the bulk electrolyte solution~\cite{delgado.av:2007}. In colloid and interface science this phenomenon has traditionally
been considered as detrimental and studied mostly in context of implications for corrections to classical theories of electrokinetic mobilities~\cite{lyklema.j:1998}. However, with the advent of nanofluidics there has been considerable interest in an unusually high conductance of electrolyte solutions confined in nanochannels, and  this subject has initiated much experimental research efforts in recent years~\cite{schoch.rb:2008}. \citet{stein.d:2004} studied ion transport in a hydrophilic silica channel of a thickness from few tens of nm to 1 $\mu$m and found a remarkable
degree of conduction at low salt concentrations that departs strongly from bulk behavior. These authors concluded that in the dilute
limit, the electrical conductances of channels saturate at a value that is independent on the salt
concentration. Later \citet{schoch.rb:2005} reported the measurements in 50 nm thick Pyrex nanoslits and found that the corresponding to a saturation plateau  conductivity augments with the surface charge density.
\citet{siria.a:2013} measured conductance of single transmembrane boron nitride nanotubes of radius ranging from 15 to 40 nm and found that the conductivity plateaus decrease with the radius.  \citet{balme.s:2015} investigated sub-10 nm
diameter hydrophobic nanopores and reported that the height of conductance plateaus is augmented, which has been interpreted in terms of slippage. During last few years several papers have been concerned with the surface conductivity of nanometric foam films and made some important remarks on the similarity to solid hydrophobic nanochannels~\cite{bonhomme.o:2015,bonhomme.o:2017}.
However, despite a rapid rise of an experimental activity  many aspects of the electro-hydrodynamics of confined electrolytes are still poorly understood theoretically or have been given insufficient attention.

 The extension of the diffuse layers is set by the nanometric Debye length, $\lambda_D \propto c_{\infty}^{-1/2}$, where $c_{\infty}$ is the concentration of the bulk electrolyte solution. Therefore, in the case of nanochannels,  EDLs of opposite walls  can occupy practically the whole channel or even strongly overlap giving  rise to physics of great complexity. Here we focus on an electrolyte solution confined in a channel whose parallel walls are separated by a distance $H$.
 Some solutions are known for parallel-plate hydrophilic channels, where no-slip boundary conditions at the walls are imposed.   \citet{levine.s:1975} addressed themselves to the case of a constant surface electrostatic potential (conducting wall) and appear to have been the first to express a surface conductivity via the  integral for the electrostatic field energy of the EDLs. They, however, have not performed detailed calculations for this energy.
  \citet{stein.d:2004} carried out such calculations assuming a constant surface charge density (insulator) and predicted the conductivity plateau at low salt, i.e. where the expected bulk conductivity nearly vanishes. There is as yet no criterion for determining \emph{a priori} whether constant charge or constant potential may be more appropriate for a particular system, and in some cases it is also possible that both surface charge density and surface electrostatic potential change with the channel thickness~\cite{ninham.bw:1971,chan.d:1975}. This phenomenon stems from the dissociation/association of surface ionizable groups and is referred to as charge regulation. It yields the self-consistent electrostatic boundary condition that differs from the two classical ones. During last years a few authors have discussed the conductivity of hydrophilic channels with strongly overlapping EDLs taking into account the charge regulation.  \citet{secchi.e:2016} reported that the conductivity exhibits a power law behavior, with an exponent close to 1/3 versus the salt concentration. \citet{biesheuvel.pm:2016} argued that this scaling is incorrect and  derived  a 1/2 power-law scaling of conductivity with $c_{\infty}$. Their results, however, apply only  when the potential is uniform across the channel and surface ionization is low, but does not describe the realistic salt concentrations. The present paper extends and generalise the earlier analysis of a hydrophilic channel, even of large surface potential and charge density, to the case of  an arbitrary salt concentration and $H$. Here we limit ourselves by the classical constant surface charge and constant surface potential electrostatic boundary conditions, which can be seen as rigorous bounds on the attainable conductivity (and its scaling with $c_{\infty}$) in the charge regulation case.

 The arguments \cite{stein.d:2004,levine.s:1975} are not complete since both electro-osmotic and electro-phoretic contributions to the
conductance could be amplified by slippage effects. Indeed, some surfaces can be slippery~\cite{vinogradova.oi:1999,vinogradova.oi:2011}, and the hydrodynamic slip length $b$ can be of the order of tens of nanometers and even more~\cite{charlaix.e:2005,vinogradova.oi:2003,joly.l:2006,vinogradova.oi:2009}. Since the efficiency of hydrodynamic slippage is determined by the ratio of $b$ to the channel thickness~\cite{vinogradova.oi:1995a}, this dramatically reduce drag and thus enhance electro-osmotic transport of ions at the nanoscale~\cite{muller.vm:1986,joly2004,silkina.ef:2019}. Besides,  at slippery surfaces adsorbed charges could be mobile, and, therefore, responding to the external electric field as discussed by \citet{maduar.sr:2015} and supported by \emph{ab initio} simulations for graphene surfaces~\cite{grosjean.b:2019}. This reduces the electro-osmotic velocity~\cite{maduar.sr:2015,silkina.ef:2019}, but could augment the electro-phoretic contribution to the channel conductance.

The quantitative understanding of impact of liquid slippage to the channel conductivity is still challenging.
Several theoretical papers have been concerned with the electrolyte conductivity in the slippery channels in the assumption that the adsorbed charges are immobile. Using a modification of the capillary pore model (space-charge theory) \citet{catalano.j:2016} proposed an expression relating the conductivity to the integral of an electrostatic potential. One of the main results of this work is that at a given surface charge the conductivity increases with the slip length,  except the low surface charge situation, where no impact of a finite slip has been found.
The authors, however, failed to propose a physical interpretation of these findings.
 \citet{bocquet.l:2010} have briefly discussed the expected shift of the conductance plateau due to a hydrodynamic slip. Applying some simple scaling arguments these authors propose that this should be $\propto b/\ell _{GC}$, where $\ell _{GC}$ is the Gouy-Chapman length, which is inversely proportional to the surface charge density~\cite{poon.w:2006}. However, they did not present any calculations illustrating or verifying their theoretical result. Similar remark applies to a paper by \citet{mouterde.t:2018} that derived scaling expressions describing a contribution of a hydrodynamic slip taking into account the mobility of adsorbed ions. Their results, however, apply  for thick channels, $H/\lambda_D \gg 1$, only. We are unaware of any attempt to quantify an effect of a mobility of adsorbed charges on a
slip-driven contribution in the case of thin channels.

Our paper is arranged as follows. In Sec.~\ref{sec:model} we define our system and summarize electrostatic relationships. Here we also introduce the notions of regimes of a thick and thin channel. Section~\ref{sec:theory} discusses an electro-hydrodynamics of a nanochannel with the focus on its conductivity. This provides insight into the convective and migration contributions to a channel conductivity, and also into the slip-driven contribution. In Sec.~\ref{sec:Hydrophilic_results} we focus attention on the conductivity of ``no-slip'' hydrophilic channels depending on electrostatic boundary conditions. Namely, equations for a conductivity in thick and thin channel regimes are derived compared with numerical results. A special attention is given to the amplification of the channel conductivity compared to a bulk value in different situations. Our treatment provides a general frameworks for interpreting experimental results and provide rigorous bounds on the conductivity for any hydrophilic channel. Hydrophobic slippery channels are analysed in detail in Sec.~\ref{sec:Hydrophobic_results}. We ascertain a dependence of slip-driven contribution on the fraction of immobile adsorbed charges at the walls that is valid for an arbitrary value of $H$, and find that it generally has a minimum, which locus depends on the value of $b/\ell _{GC}$. For large $b/\ell _{GC}$ a slip-driven contribution to the conductivity is maximized when all adsorbed ions are immobile, and can exceed the conductivity of a hydrophilic channel in a few tens of times.
  We conclude in Sec.~\ref{sec:conclusion}. Appendix~\ref{a:2} contains calculations of the mean square derivative of the electrostatic potential and of the mean osmotic pressure. The derivation of a general equation that determine the conductivity is given in Appendix~\ref{a:1}.

   \section{Summary of electrostatic relationships}\label{sec:model}

\subsection{Electrostatic model and length scales}

  \begin{figure}[h]
\begin{center}
\includegraphics[width=1\columnwidth]{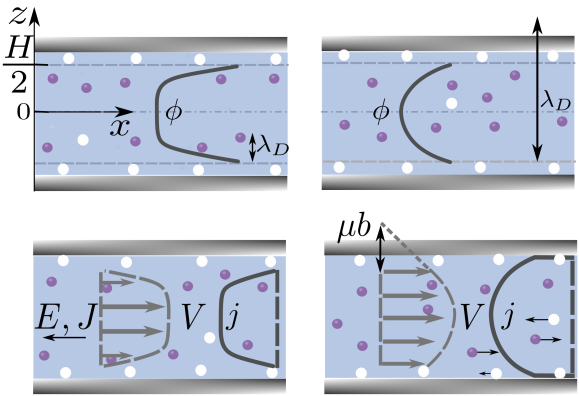}
\end{center}
\caption{
	%Schematic representation of electrostatic potential, $\phi$, electro-osmotic velocity, $v$, and local conduction current density, $j$, profiles.
	%hydrophilic (bottom left) and hydrophobic with hydrodynamic slip length $b$ (bottom right)  channels of thickness $H$.
	Schematic representation of the system that is couple with a bulk electrolyte reservoir characterized by the Debye length $\lambda_D$. The planar walls are located at $z = \pm H/2$ and separated by distance $H$. Their surface charge is created by adsorbed ions and is neutralized by the diffuse ions in the gap. Anions and cations are denoted with dark and
white circles. The distribution of an electrostatic potential $\phi(z)$ (top) is non-uniform and is qualitatively different for thick (left) and thin (right) nanochannels. An applied electric field $E$ induces an electro-osmotic velocity of a solvent $V(z)$  and a current of density $j(z)$ (bottom), which depend on the hydrodynamic slip length $b$ and on the fraction of immobile adsorbed ions $\mu$. The averaged conductivity of the nanochannel is related to its averaged current density $J$ via Ohm's law.}
\label{fig:sketch}
\end{figure}

We consider an aqueous electrolyte solution of a dynamic viscosity $\eta $
and permittivity $\varepsilon $ in a symmetric channel subject to an
electric field $E$ in the negative $x$ direction as sketched in Fig.~\ref{fig:sketch}.
The axis $z$ is defined normal to the surfaces of potential $\Phi _{s}$
and charge density $\sigma $ located at $z= \pm H/2$. Without loss of generality, the surface charges are taken as cations ($\sigma$ is positive).
For a symmetric planar channel of thickness $H$, it is enough to consider $z = H/2$ because
of the $z \leftrightarrow - z$ symmetry.
The channel is in contact
with a bulk reservoir of a 1:1 salt solution of concentration $c_{\infty}$.  Ions obey Boltzmann distribution, $c_{\pm }(z)=c_{\infty}\exp (\mp \phi (z))$, where $\phi (z)=e\Phi(z)/(k_{B}T)$ is the dimensionless electrostatic
potential, $e$ is the elementary positive charge, $k_{B}$ is the Boltzmann
constant, $T$ is a temperature of the system, and the upper (lower) sign
corresponds to the positive (negative) ions.
The Debye
screening length of an electrolyte solution, $\lambda_{D}=\left( 8\pi \ell _{B}c_{\infty}\right) ^{-1/2}$, is defined as usually with the Bjerrum
length, $\ell _{B}=\dfrac{e^{2}}{\varepsilon k_{B}T}$, where $\varepsilon $ is the permittivity of the
fluid. Note that $\ell _{B}$ of water is equal to about $0.7$ nm for room temperature. The Debye length defines an electrostatic length scale and is the measure of the thickness of the EDL. We recall that a useful formula for 1:1 electrolyte is~\cite{israelachvili.jn:2011}
\begin{equation}\label{eq:DLength}
  \lambda_D [\rm{nm}] = \frac{0.305 [\rm{nm}]}{\sqrt{c_{\infty}} [\rm{mol/L}]},
\end{equation}
so that upon increasing $c_{\infty}$ from $10^{-7}$ (in pure water, where the ionic strength is due to the dissociating H$^+$ and OH$^-$ ions) to $10^{-1}$ mol/L the screening length is reduced from about 1 $\mu$m down to ca. 1 nm.

We stress that since our discussion will be restricted to a mean-field description of the electrostatic problem and also a continuum electro-hydrodynamics, we neglect correlations and various nonidealities,  such as hydrated ion volume effects~\cite{zhu.h:2019}, dielectric mismatch~\cite{bonthuis.dj:2012}, dispersion forces between ions~\cite{ninham.bw:1997},  ion-specifity~\cite{cao.q:2018}, thermal noise~\cite{kavokin.n:2021}, which would be important for nanochannels of a few nanometers or in molecular-scale confinement. Therefore, our results apply for channels ranging from ten to several hundreds of nanometers, but in our calculations we will use only a channel of $H = 100$ nm.

The positively  charged walls attract the counter-ions (anions) and repels the co-ions (cations). In the electrolyte
solution the ions are mobile and adjust their position according to the local potential
they feel. The electrostatic potential satisfies the nonlinear Poisson-Boltzmann
equation
\begin{equation}  \label{eq:NLPB}
\phi^{\prime \prime} =\lambda _{D}^{-2}\sinh \phi
\end{equation}
where $^{\prime}$ denotes $d/d z$. We also note that as any approximation, the Poisson-Boltzmann formalism has its limits of validity, but it always describes very accurately the ionic distributions for monovalent ions in this typical concentration range~\cite{poon.w:2006}.

To integrate Eq.(\ref{eq:NLPB}) we impose two electrostatic boundary
conditions. The first condition always reflects the symmetry of the channel $%
\phi^{\prime} |_{z=0}=0$. The second condition is applied at the walls
and can be either that of a constant surface potential (conductors)
\begin{equation}
\phi |_{z=H/2}=\phi _{s},  \label{eq:bc_CP}
\end{equation}
or of a constant surface charge density (insulators)
\begin{equation}
\phi^{\prime} |_{z=H/2}=\dfrac{2}{\ell _{GC}},  \label{eq:bc_CC}
\end{equation}
where $\ell _{GC}=\dfrac{e}{2\pi \sigma \ell _{B}}$ is the Gouy-Chapman
length. These situations are referred below to as CP and CC cases.

There are a few reports of indirect surface potential measurements of CP (metal) surfaces. \citet{connor.jn:2001} tested the Poisson-Boltzmann equation directly and concluded that it gives a very accurate description of the distribution of ions in
electrolyte adjacent to a CP (metal) surface up to $\Phi_s \simeq 300$ mV ($\phi_s \simeq 12$). The latter experiments (with adsorbed mobile monolayers) produced a satisfying endorsement of equation \eqref{eq:NLPB} up to $\phi_s \simeq 20$~\cite{clasohm.ly:2006}. Several studies deduced the values of surface charge density from electrokinetic measurements. \citet{yaroshchuk.a:2009} reported  $\sigma \simeq 16-25 $ mC/m$^2$ for nanoporous track-etched membranes. The results of \citet{stein.d:2004} correspond to $\sigma \simeq 50 $ mC/m$^2$, which is comparable to reported for nanometric foam films~\cite{bonhomme.o:2015}. \citet{balme.s:2015} have interpreted their data for hydrophobic nanochannels using $\sigma \simeq 5-10 $ mC/m$^2$, which is equivalent to the range of $\ell _{GC}$ from 10 down to 5 nm. In our study we will mostly use for a typical (high) surface charge density $\sigma \simeq 25 $ mC/m$^2$, which gives $\ell _{GC} \simeq 2$ nm, but weakly charged surfaces can, of course, give much larger $\ell _{GC}$.

\begin{figure}[t]
\begin{center}
\includegraphics[width=0.99\columnwidth , trim=0.cm 0. 0.0cm
0.,clip=false]{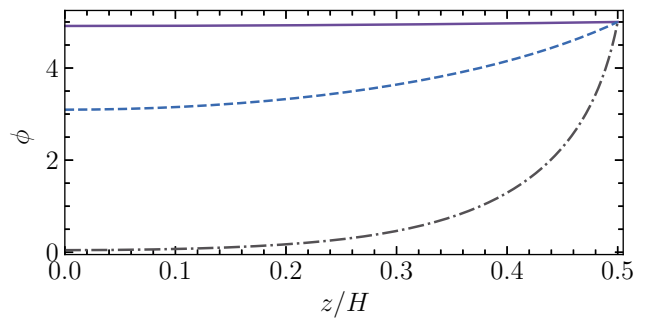}
\end{center}
\caption{Electrostatic potential computed for a CP channel of $H = 100$ nm and $\phi _{s} = 5$. From top to bottom $H/\lambda _{D} \simeq 0.1$, $1$,  and $10$. }
\label{fig:potential_profile}
\end{figure}

Figure~\ref{fig:potential_profile} shows the $\phi$-profiles in the CP channel of $H = 100$ nm calculated numerically using  $\phi _{s} = 2$ and several $\lambda_D$.  In all cases the potential is non-uniform
throughout the channel, and  the potential $\phi_m$ at the mid-plane  ($z = 0$), where the electric field vanishes, is smaller than $\phi_s$, but the form of the curves differs significantly depending on $\lambda_D$. When $H/\lambda_D \simeq 10$, the mid-plane potential practically vanish, but it is very close to $\phi_s$ for $H/\lambda_D \simeq 0.1$ indicating a strong overlap of EDLs.

Generally, the solution of Eq.\eqref{eq:NLPB} can be
obtained only numerically, but for a flat-parallel channel some exact
analytical results in the closed form can been obtained.
First integration of Eq.\eqref{eq:NLPB} from the mid-plane position to an arbitrary $z$ gives~\cite{poon.w:2006}
 \begin{equation}\label{eq:PB_out1}
\lambda_D^2 (\phi^{\prime})^2 = 2 \left[\cosh \phi - \cosh \phi_m \right],
\end{equation}
which together with \eqref{eq:bc_CC} yields a relation between the surface charge and surface and mid-plane potentials

\begin{equation}\label{eq:coshm}
\dfrac{2 \lambda_D^2}{\ell_{GC}^2} =  \cosh \phi_s - \cosh \phi_m = 2 \sinh \left(\dfrac{\phi_s + \phi_m}{2} \right)\sinh \left(\dfrac{\phi_s - \phi_m}{2} \right)
\end{equation}

Further insight can be gained by recalling that local osmotic pressure of an electrolyte solution is $P = k_B T c$, where $c (z) = c_+(z) + c_-(z)$ is the total concentration of ions at given $z$. This clarifies that $\cosh \phi$ represents a dimensionless local osmotic pressure, $p=P/2 c_{\infty} k_B T$, which takes its largest value of $\cosh \phi_s$ at $z = H/2$. Since $p(\infty) = 1$, $\cosh \phi_s - 1$ is an excess osmotic pressure
at the wall with respect to the bulk electrolyte solution.
The function $\cosh \phi$ takes its minimum at $\cosh \phi_m$. Note that $\Pi (H) = \cosh \phi_m - 1$ is an electrostatic  (dimensionless) disjoining pressure, which is the measure  of the strength of a repulsion between charged walls~\cite{derjaguin.bv:1941}.

\subsection{Thick channel regime}\label{sec:ThickCR}

If the channel is thick, $H/\lambda_D \gg 1$, the potential $\phi (z)$ decays from $\phi_s$ down to $\phi_m = 0$ and $\Pi = 0$ as seen in Fig.~\ref{fig:potential_profile}. In other words, EDLs do not overlap and
there is a bulk electrolyte solution at the mid-plane. In this regime Eq.\eqref{eq:coshm} reduces to
\begin{equation}  \label{eq:grahame}
\phi_{s} = 2\arsinh\left(\frac{\lambda _{D}}{\ell _{GC}}\right)
\end{equation}
that relates the surface potential with the charge density and is exact at any $\lambda _{D}/\ell _{GC}$.
While Eq.\eqref{eq:grahame} is identical to the Grahame equation derived for a single wall~\cite{israelachvili.jn:2011}, the boundary conditions are different.
In the present case they are dictated by symmetry, whereas in the single wall problem both $\phi$ and $\phi^{\prime}$ vanish as $z \to \infty$.

Note that Eq.\eqref{eq:grahame} suggests that $\lambda_{D}/\ell_{GC}$ represents an effective surface charge density.
For small surface charge and/or high electrolyte concentration $\lambda_{D}/\ell_{GC}$ is small, yielding $\phi_{s} \simeq 2 \ln (1 + \lambda_{D}/\ell_{GC}) \simeq 2\lambda_{D}/\ell_{GC}$, which corresponds to the linearized limit of the Poisson-Boltzmann equation \eqref{eq:NLPB}. For high surface charge and/or low salt $\lambda_{D}/\ell_{GC}$ is large, and $\phi_{s} \simeq 2 \ln (2\lambda_{D}/\ell_{GC})$, i.e. the surface potential grows weakly logarithmically with the effective surface charge.

It follows from Eq.\eqref{eq:coshm} that large $(\lambda_D/\ell_{GC})^2$ implies that $\cosh \phi_s \gg \cosh \phi_m$. In this case Eq.\eqref{eq:coshm} can be rearranged as
\begin{equation}\label{eq:graham_channel}
 \dfrac{2 \lambda_D^2}{\ell_{GC}^2} = 2 \sinh^2 \dfrac{\phi_s}{2} \left(1 - \dfrac{\Pi}{\cosh \phi_s
  - 1} \right) \simeq 2 \sinh^2 \dfrac{\phi_s}{2},
\end{equation}
indicating that Eq.\eqref{eq:grahame} should remain a sensible approximation. Thus, to use the Grahame equation it is not necessary to make assumptions about a limit of a thick channel since it also represents the rigorous asymptotic result for a non-thick channel, provided the disjoining pressure is much smaller than an excess osmotic pressure at the surfaces. Note that in this situation
\eqref{eq:PB_out1} reduces to
 \begin{equation}\label{eq:PB_out3}
\lambda_D^2 (\phi^{\prime})^2 \simeq 2 \left[\cosh \phi - 1\right],
\end{equation}
which is equivalent to
\begin{equation}\label{eq:PB_out2}
\lambda_D \phi^{\prime} \simeq 2 \sinh \left(\frac{\phi}{2}\right)
\end{equation}
The last equations are again identical to the single wall results, although EDLs do overlap. By these reasons below we refer the channel of $\Pi \ll \cosh \phi_s - 1$ to as a quasi-thick, and introduce the notion of a thick channel regime,
which integrates both truly thick and quasi-thick channels. As a side note, it has also been shown for other systems that non-thick highly charged films show the electrostatic and electro-osmotic properties of thick ones~\cite{silkina.ef:2021}.

\begin{figure}[t]
\begin{center}
\includegraphics[width=0.99\columnwidth , trim=0.cm 0. 0.0cm
0.,clip=false]{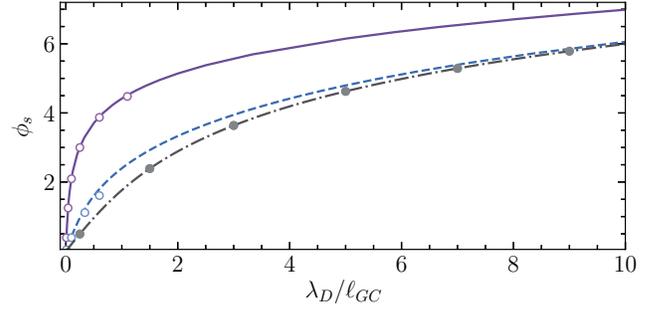}
\end{center}
\caption{Surface potential as a function $\lambda_{D}/\ell_{GC}$ calculated numerically using $H/\lambda_{D} \simeq 0.1,$ 1, and 10 (from top to bottom). Filled and open circles show calculations from Eqs.~\eqref{eq:grahame} and \eqref{eq:grahame_thin}. }
\label{fig:grahame}
\end{figure}

By setting the same values of $H/\lambda_{D}$ as in Fig.~\ref{fig:potential_profile} and  varying $\ell_{GC}$ from $1$ nm to $10$ $\mu$m it is possible to compute the curves for $\phi_s$ as a function $\lambda_{D}/\ell_{GC}$ plotted in Fig.~\ref{fig:grahame}. When $H/\lambda_D \simeq 10$, on increasing $\lambda_D/\ell_{GC}$ the surface potential
shows a weak nonlinear grow. Calculations from Eq.\eqref{eq:grahame} coincide with the numerical results,  confirming that the single wall Grahame equation holds for a thick
channel too.
When $H/\lambda_{D} \simeq 1$ the Grahame equation underestimates $\phi_s$ at small and moderate $\lambda_{D}/\ell_{GC}$, but becomes a reasonable approximation if $\lambda_{D}/\ell_{GC} \geq 5$ by confirming our above consideration. As seen in Fig.~\ref{fig:potential_profile}, at $H/\lambda_{D} \simeq 1$ the EDLs strongly overlap implying that $\Pi$ is significant, but the channel behaves effectively as thick. At first sight this result is counter-intuitive,  but  for this specific numerical example $\Pi/(\cosh \phi_s - 1) \simeq 0.1$, so that according to Eq.~\eqref{eq:graham_channel} this channel is indeed quasi-thick.
However, in the case of the channel of $H/\lambda_{D} \simeq 0.1$ in the given range of $\lambda_{D}/\ell_{GC}$ the surface potential generally significantly exceeds the predictions of Eq.\eqref{eq:grahame}, and a thick channel regime is not fulfilled unless $\lambda_{D}/\ell_{GC} \geq 50$ (not shown). For $H = 100$ nm and $\lambda_D \simeq 1$ $\mu$m this corresponds to $\ell_{GC} \leq 20$ nm. In other words, even in pure water this channel is effectively thick if its surface charge density is above  $\sigma \simeq 2.5 $ mC/m$^2$.
We return to the thin channel regime in Sec.~\ref{sec:ThinCR}.

We now define a new electrostatic length scale as
\begin{equation}  \label{eq:dukhin_length}
\ell_{Du} =  \frac{\lambda_{D}^2}{\ell_{GC}} = (8 \pi \ell_{GC} \ell_{B} c_{\infty})^{-1} = \dfrac{\sigma}{4 e c_{\infty}}
\end{equation}
A simple physical meaning of $\ell_{Du}$ is that
the layer of thickness $4 \ell_{Du}$ of the bulk solution contains an integrated charge of negatively charged counter-ions (per unit area) equal to $-\sigma$. Earlier \citet{bocquet.l:2010}
suggested that a length $\propto \lambda_{D}^2/\ell_{GC}$ should characterize the channel scale below which surface conductivity dominates over the bulk one. They termed this length the Dukhin length by analogy to the Dukhin number usually
used in colloid science.  By this reason we also refer our $\ell_{Du}$ to as the Dukhin length. Later we shell see that the conductivity of a confined electrolyte indeed depends on $\ell_{Du}$, but let us stress that $\ell_{Du}$ emerges naturally in our analysis of electrostatics.

Using $\ell _{Du}/\ell _{GC} = (\lambda_D/\ell _{GC})^2$ Eq.\eqref{eq:graham_channel} may be reexpressed as

\begin{equation}  \label{eq:grahame_Du}
\sinh \dfrac{\phi_s}{2} =  \sqrt{\frac{\ell_{Du}}{\ell_{GC}}}, \, \cosh \dfrac{\phi_s}{2} =  \sqrt{1 + \frac{\ell_{Du}}{\ell_{GC}}}
\end{equation}
These equations are valid for thick channels with any value of $\ell_{Du}/\ell_{GC}$, and also justified for quasi-thick channels, i.e. for  channels of an arbitrary thickness, provided $\ell_{Du}/\ell_{GC} \gg 1$.

Since Eq.\eqref{eq:dukhin_length} is equivalent to $\lambda_D/\ell_{GC} = \ell_{Du}/\lambda_D$, for surfaces of $\lambda_D/\ell_{GC} = 1$, $\lambda_D = \ell_{GC} = \ell_{Du}$.  In the thick channel limit $\lambda_D \ll H$, but $ \ell_{GC}$ can be smaller or larger than the channel thickness. When
 $\lambda_{D}/\ell_{GC} > 1$
\begin{equation}\label{eq:lengths}
  \ell_{GC} < \lambda_D < \ell_{Du},
\end{equation}
but $ \ell_{Du}$ can be smaller or larger than $H$. Note that $\dfrac{\ell_{Du}}{H} = \dfrac{\lambda_D}{H}\times \dfrac{\lambda_D}{\ell_{GC}}$ that increases  with a ratio of an effective surface charge to effective (electrostatic) thickness can be extremely large. Say, the values of $\lambda_{D} = 50$ nm and $\ell_{GC} = 2$ nm give $\ell_{Du} = 1.25$ $\mu$m, which is larger than any conceivable Debye length. In this case for a channel of $H = 100$ nm we obtain $\ell_{Du}/{H} \simeq 12.5$.

We now denote the average value of any function $f$ as%
\begin{equation}\label{eq:average}
\overline{f}\,=\frac{2}{H}\int\limits_{0}^{H/2}\,fdz,
\end{equation}
and in Appendix~\ref{a:2} derive expressions for the mean square derivative of the electrostatic potential $\overline{(\phi^{\prime} )^{2}}$, which is the measure of the electrostatic field energy (per unit area), and for the mean osmotic pressure $\overline{\cosh {\phi }}$ in a thick channel regime

\begin{equation}\label{eq:average_dphi}
\overline{(\phi^{\prime} )^{2}} =   \dfrac{8}{H \lambda_D} \left(\sqrt{1+\dfrac{\ell _{Du}}{\ell _{GC}}} - 1\right),
\end{equation}
\begin{equation}\label{eq:average_cosh}
\overline{\cosh\phi} =  1 + \dfrac{4\lambda_D}{H }\left( \sqrt{1+\dfrac{\ell _{Du}}{\ell _{GC}}}-1\right)
\end{equation}

\subsection{Thin channel regime}\label{sec:ThinCR}

We now turn to the thin channel limit, $H/\lambda _{D}\ll 1$, where the EDLs strongly overlap (see Fig.~\ref{fig:potential_profile}). If we make the addition assumption that $\lambda_D/\ell_{GC}$ is small enough, the disjoining pressure $\Pi$ becomes very close to the excess osmotic pressure at the wall. We will term this situation as a thin channel regime, which is a more narrow notion compared to a thin channel limit. It is well seen in Fig.~\ref{fig:grahame} that in the thin channel regime $\phi_s$ rises much more rapidly (and nonlinearly) with $\lambda_D/\ell_{GC}$ then prescribed by the Graham equation, and that $\phi_s$ becomes quite large despite $\lambda_D/\ell_{GC} \le 1$ (weakly charged surfaces). We remark that for $\lambda_{D}/\ell_{GC} < 1$ the electrostatic lengths are organised in the order
\begin{equation}\label{eq:lengths_small}
 \ell_{Du} < \lambda_D < \ell_{GC},
\end{equation}
i.e. in a thin channel regime $\ell_{Du}$ is the smallest electrostatic length of the problem, but $H$ can be smaller or larger than $\ell_{Du}$.

\citet{silkina.ef:2019} have recently shown that in this regime the distribution of a potential in
the channel is
\begin{equation}  \label{eq:series}
\phi (z)\simeq\phi _{s}+\dfrac{\sinh \phi _{s}}{2 \lambda _{D}^{2}}\left(
z^{2}-\dfrac{H^2}{4}\right)
\end{equation}
leading to
\begin{equation}  \label{eq:grahame_thin}
\sinh \phi_s \simeq  \frac{4 \ell_{Du}}{ H}, \, \cosh \phi_s \simeq \sqrt{ 1 + \left( \dfrac{4\ell_{Du}}{H} \right)^2 }
\end{equation}

The surface potential calculated from \eqref{eq:grahame_thin} is included in Fig.~\ref{fig:grahame} (shown by the open circles) and we see that the fits are
quite good both for truly thin, $H/\lambda _{D}\ll 1$, and quasi-thin, $H/\lambda _{D} \simeq 1$ channels. Therefore, the condition of a thin channel regime can be relaxed to  $H/\lambda _{D}\le 1$ and $\lambda_D/\ell_{GC} \le 1$.

It follows from \eqref{eq:series} and \eqref{eq:grahame_thin} that the potential at the mid-plane of a thin channel is given by
\begin{equation}
	\phi_m \simeq \phi_s - \dfrac{H}{2 \ell_{GC}},
		\end{equation}%
where $H/ \ell_{GC}$ is much smaller than $\phi_s$.

In the thick channel regime the impact of $\Pi$ can safely be neglected as demonstrated in Sec.~\ref{sec:ThickCR}, but for a regime of a thin channel Eqs.\eqref{eq:coshm} and \eqref{eq:grahame_thin} give
\begin{equation}\label{eq:Pi_thin}
  \Pi \simeq \sqrt{1 + \left( \dfrac{4 \ell_{Du}}{H}\right)^2} - 1 - \dfrac{2 \ell_{Du}}{\ell_{GC}}
\end{equation}
The last equation shows that the electrostatic disjoining pressure is equal to the excess osmotic pressure at the wall with a leading (small) correction term $2 \ell_{Du}/\ell_{GC}$.
When $H$ is the smallest length scale in the system, $\Pi \simeq 4 \ell_{Du}/{H}$.

Finally, in the thin channel regime $\overline{(\phi^{\prime} )^{2}}$ and $\overline{\cosh\phi}$ can be approximated by (see Appendix~\ref{a:2} for a derivation)

\begin{equation}\label{eq:series_der_thin}
\overline{(\phi^{\prime} )^{2}} \simeq  \dfrac{4}{3 \ell_{GC}^{2}}
\end{equation}
and
\begin{equation}\label{eq:average_cosh_thin}
\overline{\cosh\phi} \simeq \sqrt{ 1 + \left( \dfrac{4\ell_{Du}}{H} \right)^2 }- \dfrac{4 \ell_{Du}}{3 \ell_{GC}}
\end{equation}

\subsection{Constant charge vs. constant potential}\label{sec:CCCP}

Finally, we stress that the derived above equations can be used both for CC and CP cases.

For a CC channel $\ell_{GC}$ is fixed, and does not depend on the concentration of salt and on the channel thickness. Therefore, $\ell_{Du} \propto c_{\infty}^{-1}$ irrespectively of $H$. Naturally, the same scaling law holds for $\ell_{Du}/H$ and $\ell_{Du}/\ell_{GC}$. It follows from Eq.\eqref{eq:Pi_thin} that if $\ell_{Du}/H \geq 1$, in the thin channel regime the disjoining pressure is inversely proportional to the thickness and salt concentration, $\Pi \propto (c_{\infty} H)^{-1}$.

\begin{table}
  \centering
    \renewcommand{\baselinestretch}{2}\normalsize
  \caption{The Dukhin and Gouy-Chapman lengths and useful ratios expressed via. the surface potentials.}
        \label{table:CPCC}
  \begin{tabular}{|c|c|}
  %\hline
  \multicolumn{2}{c}{} \\
  \hline
   Thick channel regime & Thin channel regime \\
    \hline
    $ \ell_{Du} =  \lambda_D \sinh \dfrac{\phi _{s}}{2}$  &  $\ell_{Du} \simeq \dfrac{H \sinh \phi_s}{4} $  \\
    \hline
     $\ell_{GC} = \dfrac{\lambda_D}{\sinh \dfrac{\phi _{s}}{2}}$  & $\ell_{GC} \simeq \dfrac{4 \lambda_D^2}{H \sinh \phi _{s}} $ \\
      \hline
       $ \dfrac{\ell_{Du}}{\ell_{GC}} =   \sinh^2 \dfrac{\phi _{s}}{2}$  &  $\dfrac{\ell_{Du}}{\ell_{GC}} \simeq \left(\dfrac{H}{\lambda_D}\right)^2 \dfrac{\sinh^2 \phi_s}{16 } $  \\
    \hline
       $ \dfrac{\ell_{Du}}{H} =  \dfrac{\lambda_D}{H} \sinh \dfrac{\phi _{s}}{2}$  &  $\dfrac{\ell_{Du}}{H} \simeq \dfrac{\sinh \phi_s}{4} $  \\
    \hline
        \end{tabular}
        \end{table}

In the CP channel $\phi_s$ is fixed and, depending on the regime, expressions for $\ell_{Du}$ follow directly from Eqs.\eqref{eq:grahame_Du} or \eqref{eq:grahame_thin}.
Standard manipulations yield the expressions for  $\ell_{Du}$ and $\ell_{GC}$ of a CP channel, which are summarised in Table~\ref{table:CPCC}. Also included are the useful ratios of the lengths that appeared in the above analysis. An important difference of CP channels is that the expressions for $\ell_{Du}$ and $\ell_{GC}$ are different in thick and thin channel  regimes.

For the regime of a thick CP channel
$\ell_{Du}\propto c_{\infty}^{-1/2}$. The same scaling law of $\ell_{GC}$, $\ell_{Du}/H$, and $\ell_{GC}/H$ with salt concentration is hold, but $\ell_{Du}/\ell_{GC} = (\lambda_D/\ell_{GC})^2 $ is an independent on $c_{\infty}$ constant set solely by $\phi_s$. We recall that in this regime $\Pi$ can safely be excluded from the analysis, even when it is not small.

In the regime of a thin CP channel $\ell_{Du}$ does not depend on salt, and $\ell_{Du}/H$ is a constant that reflects solely the surface potential. We also stress that $\ell_{GC} \propto c_{\infty}^{-1}$, and, therefore, $\ell_{Du}/\ell_{GC} \propto c_{\infty}$. The disjoining pressure is given by
\begin{equation}\label{eq:Pi_thin_CP}
  \Pi \simeq \cosh \phi_s - 1 - \left(\dfrac{H}{\lambda_D}\right)^2 \dfrac{\sinh^2 \phi_s }{8},
\end{equation}
i.e. $\Pi$ tends to a constant value expected when $H \to 0$ with a correction $\propto c_{\infty} H^2$. Note that Eq.\eqref{eq:Pi_thin_CP} is identical to derived by \citet{markovich.t:2016} for the CP case.

\section{General considerations}\label{sec:theory}

The system subjects to a tangential electric field $E$ [V/m] that induces an electro-osmotic flow of velocity $V(z)$ [m/s] that satisfies Stokes' equations with an electrostatic body force:
\begin{equation}  \label{eq:Stokes}
  v^{\prime \prime} = \phi^{\prime \prime},
\end{equation}
where $v(z) = \dfrac{4\pi \eta
\ell _{B}}{e E} V(z)$ is the dimensionless fluid velocity.

The fluid velocity at $z=H/2$ satisfies~\cite{maduar.sr:2015,silkina.ef:2019}
\begin{equation}
v|_{z=H/2}=b\left( - \phi^{\prime}|_{z=H/2}+\dfrac{2(1-\mu) }{\ell _{GC}}\right) = -\dfrac{2\mu b }{\ell _{GC}}  \label{eq:bc_Stokes2},
\end{equation}%
where to obtain second relation we used Eq.\eqref{eq:bc_CC}. The parameter $\mu$ in \eqref{eq:bc_Stokes2} is the fraction of immobile surface changes that can vary from 0 for fully mobile charges to 1 in the case when all adsorbed ions are fixed.
Eq.\eqref{eq:bc_Stokes2}
implies that $v(H/2) \propto -\mu b \sigma$. In particular, we see that the velocity at the wall reaches its maximal possible value, $v(H/2) = - b/\ell _{GC}$,  when surface ions are immobile. We also see, that if surface ions are fully mobile, $v(H/2)$ vanishes even when hydrodynamic slip length $b$ is large, so that an electro-osmotic flow near such surfaces would be identical to that near (no-slip) hydrophilic walls.

The second hydrodynamic boundary condition is implied by symmetry:
\begin{equation}
v^{\prime}|_{z=0}=0.  \label{eq:Stokes_sym}
\end{equation}

Performing the integration in Eq.\eqref{eq:Stokes} with prescribed boundary conditions (\ref{eq:bc_Stokes2}) and (\ref{eq:Stokes_sym}) yields~\cite{silkina.ef:2019}
\begin{equation}
v(z)=\phi (z)-\phi _{s} - \frac{2 \mu b}{\ell _{GC}}
\label{eq:Stokes_solution}
\end{equation}%

An applied electric field also induce an electric current in the channel. The local current density $j(z)$ is, of course, not uniform  and depends on $z$. It is convenient to introduce an averaged density $J = \overline{j}$ [A/m$^2$]. The averaged
channel conductivity $K$ [S/m] is then related to $J$ by $K = J/E$ (Ohm's law) and includes two contributions. Namely, a convective contribution due to an electro-osmotic flow of a solvent (of a velocity $v$) and a migration contribution caused by an electro-phoretic motion of ions with respect to the solvent. To calculate $J$ we assume a weak field, so that in steady state $\phi (z)$ is independent of the fluid flow.
Note that in experiment one measures a conductance, $G$ [S]. For a flat-parallel channel $G = K \dfrac{w H}{L}$ , where $L$ is the length of the channel and $w$ is its width, so that the cross-sectional area is given by $w H$.

 The local current density of thermal ions in confined electrolyte is $j_{+}+j_{-}$, where $j_{\pm }=\pm ec_{\pm }(V \pm m_{i}E)$. The first term is associated with the convective contribution, i.e. with the transport of ions by an electro-osmotic flow of velocity $v$ given by Eq.\eqref{eq:Stokes_solution}. The second term represents a migration contribution caused by the (electro-phoretic) movement of ions with respect to the solvent.
The migration term depends on the mobility of ions $m_{i}$ that can be expressed as $m_{i}=e/(6\pi \eta \mathcal{R_i})$. Here for simplicity we assume that the radii of positive and negative ions are equal to  $\mathcal{R}$, so that they are of the same mobility. In our calculations below we will use $\mathcal{R} = 0.3$ nm. Now, by adding the contribution of adsorbed ions, $j_{\sigma}$, to conductivity  we can formulate the expression for a mean current density as
\begin{equation}
J =  \overline{j_{+}}+\overline{j_{-}} + \dfrac{2 j_{\sigma}}{H},
\label{eq:J}
\end{equation}
The detailed derivations of the expressions of these contributions is given in Appendix~\ref{a:1}. There to calculate $ j_{\sigma}$ we make an additional assumption that the mobility of adsorbed surface charges is the same as that for diffuse ions in electrolyte solution. This is, of course, the largest possible mobility at surfaces that would allow us to maximize the effect of mobile adsorbed ions on the total conductivity, but note that such a situation is more than realistic. As discussed by \citet{lyklema.j:1998} the lateral mobilities of monovalent ions in the adsorbed  layer are not much lower than those in bulk and are very often of the same order of magnitude.

\begin{figure}[h]
\begin{center}
\includegraphics[width=1\columnwidth]{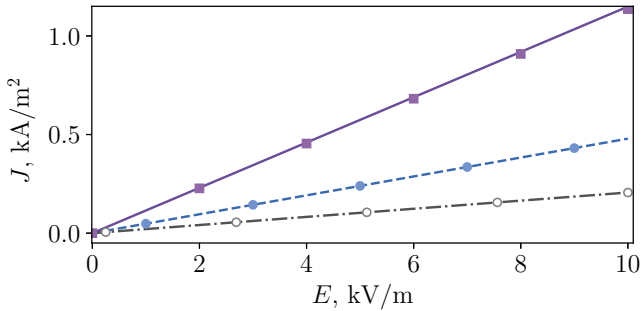}
\end{center}
\caption{Current density $J$ vs. electric field $E$ computed for channels of $H = 100$ nm by imposing different electrostatic and electro-hydrodynamic boundary conditions.  The solid line shows numerical results for a hydrophobic CP channel obtained using $\phi_{s} = 4$, $\mu = 0.5$, $b = 100$ nm, and $c_{\infty} = 10^{-3}$ mol/L. Alternate lines are computed for hydrophilic CC channels of $\ell_{GC} = 1$ nm using $c_{\infty} = 10^{-3}$ mol/L  (dashed) and $c_{\infty} = 5 \times 10^{-6}$ mol/L (dash-dotted). Filled and open circles are calculations from Eqs.\eqref{eq:hydrophilic_thick} and \eqref{eq:hydrophilic_thick3}, correspondingly. Squares show predictions of Eq.\eqref{eq:sc}.}
\label{fig:conduction_current}
\end{figure}

Figure~\ref{fig:conduction_current} shows a typical current-voltage response ($J-E$) of several nanochannels of the same thickness, but different surface properties, computed for two concentrations of salt, $c_{\infty} = 10^{-3}$ mol/L and $5\times10^{-6}$ mol/L. The calculations for CC channel are made using $\ell_{GC} = 1$ nm and with no slippage at the surfaces.  The slope of the $J-E$ straight line is invoked to find the conductivity $K$ that depends on $c_{\infty}$, and our numerical examples show that $K$ is larger for $c_{\infty} = 10^{-3}$ mol/L. The CC data are compared with another numerical calculation using $c_{\infty} = 10^{-3}$ mol/L, but made for a CP channel of $\phi_{s} = 4$ in which a slip length of 100 nm and $\mu = 0.5$ are incorporated. The computed data show larger $K$ than in the case of ``no-slip'' CC channel indicating that $K$ depends not solely on the salt concentration, but also on the surface properties reflected by electrostatic and electro-hydrodynamic boundary conditions. To explain these results a quantitative theory of the channel conductivity is required. The theory must be able to account for electrostatic and wetting properties of the surfaces and provide a realistic description of the bulk configuration. We develop a suitable theory and return later to the fit of the curves shown in Fig.~\ref{fig:conduction_current}.

It is convenient to divide the mean conductivity of the channel into a ``no-slip'' conductivity $K_0$ expected for hydrophilic channels, and slip-driven contribution $\Delta K$:

\begin{equation}
K = K_0 + \Delta K
\label{eq:M22}\end{equation}

A derivation of a general equation that determines $K$ is given in Appendix~\ref{a:1} and yields the following expressions for a  conductivity of an electrolyte solution in the bulk
\begin{equation}
	K^{\infty }=\frac{e^{2}}{24\pi ^{2}\eta \ell _{B}\mathcal{R}\lambda _{D}^{2}} =\frac{e^{2} c_{\infty}}{3 \pi \eta \mathcal{R}},%
	 \label{eq:ic0_thick}
	\end{equation}%
and inside the hydrophilic channel of an arbitrary thickness
\begin{equation}\label{eq:K0}
K_0 = K^{\infty }\displaystyle \left[ \dfrac{3 \lambda _{D}^{2}\mathcal{R}\overline{(\phi^{\prime} )^{2}}}{2\ell _{B}}+ \overline{\cosh {\phi }}\right],
\end{equation}
The first and second terms are associated with the convective and migration contributions, correspondingly.
Eq.\eqref{eq:K0} is identical to the \citet{levine.s:1975} integral formula, but here $K_0$ is related to the mean values of $\overline{(\phi^{\prime} )^{2}}$ and $\overline{\cosh {\phi }}$.
To determine $K_0$ given by Eq.\eqref{eq:K0} we have to substitute the relevant expressions for $\overline{(\phi^{\prime} )^{2}}$ and $\overline{\cosh {\phi }}$, which have different forms in the regimes of thick and thin channels as discussed in Sec.\ref{sec:ThickCR} and \ref{sec:ThinCR}. In Sec.~\ref{sec:Hydrophilic_results} we will see that this variety of possible situations gives rise to a rich diversity in the conductivity regimes of even canonical hydrophilic channels.

The slip-driven contribution $\Delta K$ is of the same form for a channel of any thickness and is given by (see Appendix~\ref{a:1})

\begin{equation}\label{eq:slip correction}
  \Delta K = K^{\infty }\frac{4 \ell_{Du}}{H} \left( \frac{3 \mu ^{2}b}{\ell
_{GC}}\dfrac{\mathcal{R}}{\ell _{B}}+1-\mu \right)
\end{equation}
The first term is associated with an additional convective conductivity due to a hydrodynamic slip. The  proportional to $(1 - \mu)$ term represents a migration contribution of mobile adsorbed ions. Recently \citet{mouterde.t:2018} derived an equivalent expression for the slip-induced surface conductivity by assuming $H \gg \lambda_D$. Our analysis
clarifies that Eq.\eqref{eq:slip correction}  constitutes a rigorous result for $\Delta K$ of a channel of any $H$.
This allows us to make contact with the semi-analytical calculations  of \citet{catalano.j:2016} mentioned above (in Sec.~\ref{sec:intro}). Eq.\eqref{eq:slip correction} provides an explanation of their results that correspond to $\mu = 1$. Indeed, when $\mu \to 1$, only a convective contriution plays a role. For very weakly charged surfaces, i.e. at $b/\ell
_{GC} \to 0$, this convective term  becomes negligibly small leading to $\Delta K \to 0$. Thus, in this limit we recover a ``no-slip'' solution that has been found before numerically~\cite{catalano.j:2016}.
When $b/\ell
_{GC}$ is finite the function $\Delta K$ is positive definite for any $\mu$ (from 0 to 1).

Differentiating Eq.\eqref{eq:slip correction} twice with respect to $\mu$ we conclude that $d^2 (\Delta K) / d \mu^2$ is always positive, and $\Delta K$
takes its minimum value of
\begin{equation}\label{eq:slip correction_min}
  \Delta K = K^{\infty }\frac{4 \ell_{Du}}{H} \left(1-\frac{\ell
_{GC}\ell
_{B} }{9 b \mathcal{R}}\right)
\end{equation}
at $\mu = \ell_{GC}/ b \times \ell_{B}/6 \mathcal{R}$. This implies that if we keep $b/\ell_{GC}$ fixed, but increases $\mu$, $\Delta K$ reduces until it reaches its minimum defined by Eq.\eqref{eq:slip correction_min}. On increasing $\mu$ further $\Delta K$ should demonstrate a  fast parabolic growth. Note that for $b/\ell_{GC} < \ell_{B}/6 \mathcal{R}$ the minimum of $ \Delta K$ occurs outside of the possible range of $\mu$.
In this situation the value of $\Delta K$  reduces with $\mu$   and takes its smallest possible value at $\mu = 1$.
We will return to the importance of $\Delta K$ and physics underlying its behavior in Sec.~\ref{sec:Hydrophobic_results}.

\section{Hydrophilic channels}\label{sec:Hydrophilic_results}

In the thick channel regime $\overline{(\phi^{\prime} )^{2}}$ and $\overline{\cosh {\phi }}$ are
defined by Eqs.\eqref{eq:average_dphi} and \eqref{eq:average_cosh}.
The conductivity of a hydrophilic channel in this regime is then given by

\begin{equation}\label{eq:hydrophilic_thick}
K_0 \simeq  K^{\infty } \left[ 1 + \dfrac{4\lambda_D}{H }\left( \sqrt{1+\dfrac{\ell _{Du}}{\ell _{GC}}}-1\right)\left(1 + \dfrac{3 \mathcal{R}}{ \ell _{B}} \right) \right]
\end{equation}
 The second term in \eqref{eq:hydrophilic_thick} represents the surface conductivity that vanishes in the limits of $H \to \infty$ (a single wall) or $\ell _{Du}/\ell _{GC} \to 0$ (uncharged surfaces). Note that Eq.(15) by \citet{mouterde.t:2018} for the surface conductivity (given without a derivation for the limit of $H/\lambda_D \gg 1$) can be rearranged to the second term of Eq.\eqref{eq:hydrophilic_thick}. We stress that for a truly thick channel Eq.\eqref{eq:hydrophilic_thick} becomes exact and is applicable for any $\ell _{Du}/\ell _{GC}$, but it should also constitute very good an approximation for highly charged non-thick channels.

For weakly charged surfaces (small $\ell _{Du}/\ell _{GC}$) and $H/\lambda_D \gg 1$, Eq.\eqref{eq:hydrophilic_thick} reduces to
\begin{equation}\label{eq:weak}
K_0 \simeq  K^{\infty } \left[ 1 + \dfrac{2 \lambda_D\ell _{Du}}{H \ell _{GC}}\left(1 + \dfrac{3 \mathcal{R}}{ \ell _{B}} \right) \right]
\end{equation}
that suggests that the surface conductivity constitutes a small correction to the bulk one and, therefore, can safely be neglected.

If $\ell _{Du}/\ell _{GC}$ is large, i.e. when a thick channel regime also includes the quasi-thick channels as discussed in Sec.~\ref{sec:ThickCR},
Eq.\eqref{eq:hydrophilic_thick}  can be simplified to give
\begin{equation}\label{eq:hydrophilic_thick2}
K_0 \simeq K^{\infty } \left[ 1 + \dfrac{4 \ell _{Du}}{H }\left(1 + \dfrac{3 \mathcal{R}}{ \ell _{B}} \right) \right],
\end{equation}
which is valid for any $ \ell _{Du}/H$. Estimating the orders of magnitude we conclude that the surface conductivity dominates when $4 \ell _{Du}/H \geq 1$. We recall that the bulk electrolyte layer of thickness $4 \ell _{Du}$ of  contains an integrated charge
of counter-ions (per unit area), which is exactly equal to  the surface charge density taken with the opposite sign. Thus this result is not surprising, although it could not be predicted \emph{a priori}. However, to safely neglect the bulk contribution to $K_0$ we should require $\ell _{Du}/H \geq 1$. In other words, we might argue that when $\ell _{Du}$ exceeds $H$, a sensible approximation for $K_0$ should be
\begin{equation}\label{eq:hydrophilic_thick3}
K_0 \simeq K^{\infty }  \dfrac{4 \ell _{Du}}{ H }\left(1 + \dfrac{3 \mathcal{R}}{ \ell _{B}} \right)
\end{equation}
Thus, even moderate $ \ell _{Du}/H$ greatly enhances the surface conductivity of a hydrophilic channel if and only if $\lambda_D/\ell _{GC}$ is large. Clearly, large $ \ell _{Du}/H$ should lead to a huge amplification of $K_0$.

In the thin channel regime, which implies that walls are necessarily relatively weakly charged (see Sec.~\ref{sec:ThinCR}), $\overline{(\phi^{\prime} )^{2}}$ and $\overline{\cosh {\phi }}$ are
given by Eqs.\eqref{eq:series_der_thin} and \eqref{eq:average_cosh_thin}. Substitung them to \eqref{eq:K0} we derive

\begin{equation}\label{eq:hydrophilic_thin}
K_0 \simeq K^{\infty } \left[\sqrt{ 1 + \left( \dfrac{4\ell_{Du}}{H} \right)^2 } - \dfrac{4 \ell_{Du}}{3 \ell _{GC}} \left(1 - \dfrac{3 \mathcal{R}}{2 \ell_{B}}, \right)\right]
\end{equation}
where the first term is of the leading-order.
Clearly, the significant deviations from the bulk conductivity are expected only when $\ell_{Du}/H \geq 1$. In this case Eq.\eqref{eq:hydrophilic_thin} reduces to
\begin{equation}\label{eq:hydrophilic_thin2}
K_0 \simeq K^{\infty }  \dfrac{4\ell_{Du}}{H} \left[ 1 - \dfrac{H}{3 \ell _{GC}} \left(1  - \dfrac{3 \mathcal{R}}{2 \ell_{B}} \right)\right] \simeq K^{\infty } \dfrac{4\ell_{Du}}{H}
\end{equation}
Thus, when $H$ becomes the smallest length scale of the problem, the conductivity is bounded by the value of the disjoining pressure and can be enhanced in nearly $\Pi$ times compared to the bulk case.  Already at this point one can conclude that for a very thin CC channel the conductivity amplification, $K_0/K^{\infty }$, can be very large for a few nm channels. This follows from the fact that the electrostatic disjoining pressure scales as $\Pi \propto (c_{\infty} H)^{-1}$, i.e. diverges at $H \to 0$. However, for a vanishing thickness the disjoining pressure in the CP channel tends to a constant value as follows from Eq.\eqref{eq:Pi_thin_CP}, so does $K_0/K^{\infty }$.

\begin{figure}[h]
\begin{center}
\includegraphics[width=1\columnwidth]{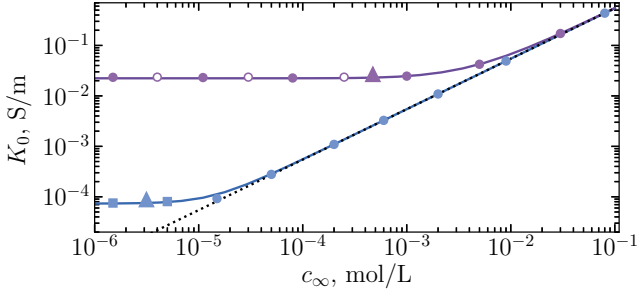}
\end{center}
\caption{$K_0$ as a function of
 $c_{\infty}$ computed for CC channels of $H = 100$ nm (solid curves) using $\ell_{GC} = 2$ and 300 nm (from top to bottom). Dotted line shows $K^{\infty }$  calculated from Eq.\eqref{eq:ic0_thick}. Filled circles show predictions of Eq.\eqref{eq:hydrophilic_thick}, open circles show calculations from Eq.\eqref{eq:hydrophilic_thick3}. Filled squares are obtained from Eq.~\eqref{eq:hydrophilic_thin}.  The big triangles mark the points of $\ell_{Du}/{H} = 1$.}
\label{fig:conductivity}
\end{figure}

If for a channel of $H = 100$ nm we keep $\ell_{GC}$  fixed, on varying $c_{\infty}$ it is possible to obtain the conductivity curves displayed in Fig.~\ref{fig:conductivity}. For this specimen examples we use $\ell_{GC} = 2$ nm ($\sigma \simeq 18 $ mC/m$^2$) and 300 nm ($\sigma \simeq 0.12 $ mC/m$^2$). In both cases $K_0$ remains constant at sufficiently small concentrations. For larger $c_{\infty}$ the $K_0$-curves converge to the bulk conductivity  calculated from Eq.\eqref{eq:ic0_thick} and increase linearly with $c_{\infty}$. The big triangles correspond to $\ell_{Du}/{H} = 1$ and we see that they provide a good sense of the transition from the saturation (plateau) regions to the bulk branch of the conductivity curves. The plateau occurs when $\ell_{Du}/{H} \geq 1$. For an upper curve of $\ell_{GC} = 2$ nm $\lambda_D/\ell_{GC}$ is large, indicating a thick channel regime even when EDLs strongly overlap. The theoretical curve calculated from Eq.\eqref{eq:hydrophilic_thick} is also included in Fig.~\ref{fig:conductivity}. We see that the fits are quite good for all $c_{\infty}$.
The plateau branch, where $\ell_{Du}/{H} \geq 1$, is reasonably well described by a more elegant theoretical result, Eq.\eqref{eq:hydrophilic_thick3}.
The nature of the plateau is apparent now. An accuracy of
Eq.\eqref{eq:hydrophilic_thick3} implies that $K_0 \propto K^{\infty } \ell _{Du}$ that does not depend on $c_{\infty}$ for a CC channel. We remark that for a conductivity plateau the conductance $G \propto K_0 H$ does not depend on $H$. This is exactly what has been observed by  \citet{stein.d:2004}. These authors also found experimentally that the hight of the conductivity  plateau increases with surface charge density. Our theoretical results agree well with this conclusion too. Indeed, the curve of  $\ell_{GC} = 300$ nm is located much lower than that for a highly charged channel. For  very dilute solutions, $\ell_{Du}/{H} \geq 1$, this curve corresponds to a thin channel regime, and the plateau is well described by Eq.\eqref{eq:hydrophilic_thin2}, but for larger $c_{\infty}$ the condition of a thick channel regime is fulfilled and Eq.\eqref{eq:weak} becomes valid.

\begin{figure}[h]
\begin{center}
\includegraphics[width=1\columnwidth]{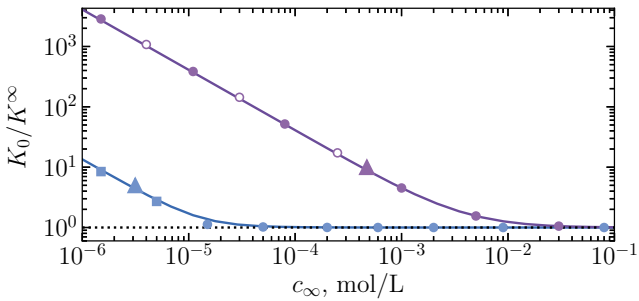}
\end{center}
\caption{The data sets of Fig.~\ref{fig:conductivity} reproduced, and plotted as $K_{0}/K^{\infty}$ vs. $c_{\infty}$.}
\label{fig:conduction}
\end{figure}

To examine the significance of the surface conductivity contribution  (in respect to the bulk one) more closely, the data sets from Fig.~\ref{fig:conductivity} are reproduced in Fig.~\ref{fig:conduction}, but plotted as $K_{0}/K^{\infty}$ vs. $c_{\infty}$. On reducing $c_{\infty}$ (increasing $\lambda_D$ and $\ell_{Du}$) the relative contribution of the surface conductivity weakly increases until $\ell_{Du}/{H} \simeq 1$. On decreasing $c_{\infty}$ further $K_{0}/K^{\infty}$ grows linearly in this log-log plot indicating that $K_{0}/K^{\infty} \propto c_{\infty}^{-1}$.  The surface conductivity can become very large (a few order of magnitude larger than $K^{\infty}$),  for highly charged quasi-thick channels, but even for an extremely weakly charged walls  one can obtain an order of magnitude enhancement,  provided
that an electrolyte solution is very dilute, i.e. in the thin channel regime.

\begin{figure}[h]
\begin{center}
\includegraphics[width=1\columnwidth]{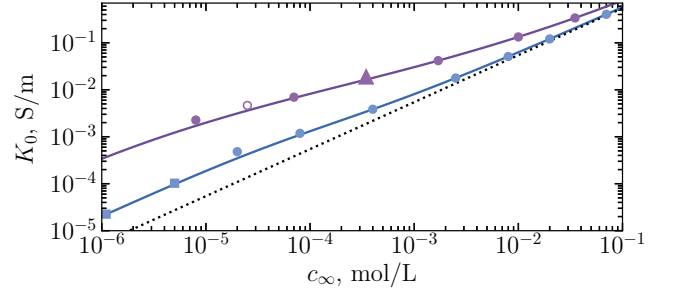}
\end{center}
\caption{$K_0$ as a function of
 $c_{\infty}$ computed for CP channels of $H = 100$ nm (solid curves) using $\phi_{s} = 5$ and 2 (from top to bottom). Dotted line shows $K^{\infty }$  calculated from Eq.\eqref{eq:ic0_thick}. Filled circles show predictions of Eq.\eqref{eq:hydrophilic_thick}, open circles show calculations from Eq.\eqref{eq:hydrophilic_thick3}. Filled squares are obtained from Eq.~\eqref{eq:hydrophilic_thin}. The big triangles mark the points of $\ell_{Du}/{H} = 1$.}
\label{fig:conductivity_CP}
\end{figure}

If we keep $\phi_s$ fixed and vary $c_{\infty}$, we move to the situation shown in Fig.~\ref{fig:conductivity_CP}. For this numerical example we use $\phi_s = 5$ ($\Phi_s \simeq 125$ mV) and 2 ($\simeq 50$ mV). Note that for a channel of $\phi_s = 5$ the thick channel regime is expected when $c_{\infty} \geq 10^{-5}$ mol/L. At smaller concentrations this channel falls to the intermediate regime (between thick and thin), and even at $c_{\infty} = 10^{-6}$ mol/L the value of $\Pi/(\cosh \phi_s - 1) \simeq 0.6$ is still not large enough to justify thin channel approximations. The value of $\ell_{Du}/{H} = 1$ corresponds to $c_{\infty} \simeq 3.4\times 10^{-4}$ mol/L, i.e. this is located at the branch of the curve, where the thick channel regime is valid. The channel of $\phi_s = 2$ is thin when $c_{\infty} \leq 2 \times 10^{-6}$ mol/L that results in constant $\ell_{Du}/{H} \simeq 0.91$ on the whole branch of a thin channel regime (see Sec.~\ref{sec:CCCP}). Since $\ell_{Du}$  can only decrease with salt, but never increases, this implies that with this value of $\phi_s$ the value of $\ell_{Du}/{H} = 1$ is never reached. In other words, the surface conductivity of this channel never  exceeds the bulk one significantly. Finally, we note that when $\phi_s = 2$, the thick channel regime occurs at $c_{\infty} \geq 10^{-4}$ mol/L.
Figure~\ref{fig:conductivity_CP} shows that both for $\phi_s = 5$ and 2 the conductivity of a hydrophilic channel, $K_0$, strictly monotonously increases with $c_{\infty}$, and the saturation plateau does never occur. The deviations from the bulk conductivity are larger for a larger $\phi_s$, and for $\phi_s = 5$ are discernible even for concentrated solutions, but at $\phi_s = 2$ the conductivity curve converges to $K_{\infty}$ already when $c_{\infty} \simeq 10^{-2}$ mol/L. Also included in Fig.~\ref{fig:conductivity_CP} are the theoretical calculations from Eqs.\eqref{eq:hydrophilic_thick},  \eqref{eq:hydrophilic_thick3}, and \eqref{eq:hydrophilic_thin}, where we have substituted $\ell_{Du}$ and $\ell_{GC}$ by the relevant expressions from Table~\ref{table:CPCC}. We see that these approximations fit quite good the appropriate branches of the conductivity curves.

\begin{figure}[h]
\begin{center}
\includegraphics[width=1\columnwidth]{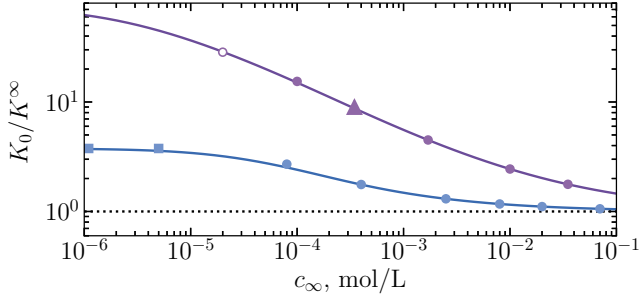}
\end{center}
\caption{The data sets of Fig.~\ref{fig:conductivity_CP} reproduced, and plotted as $K_{0}/K^{\infty}$ vs. $c_{\infty}$.}
\label{fig:conduction_CP}
\end{figure}

To demonstrate the salt dependence and magnitude of the conductivity enhancement in the CP case, in Fig.~\ref{fig:conduction_CP} the numerical and theoretical data are reproduced from Fig.~\ref{fig:conductivity_CP}, but plotted for $K_{0}/K^{\infty}$. At low salt we observe a saturation of $K_{0}/K^{\infty}$ for a curve of $\phi_s = 2$. This is an indication of a thin channel regime and the height of this plateau is roughly equal to $\sinh \phi_s \simeq 3.6$, i.e. the conductivity of this channel is only a few times higher than the bulk one (which is extremely small at low salt). It follows from Eq.\eqref{eq:hydrophilic_thin} that the amplification of $K_0$ cannot exceed $\cosh \phi_s$, and for salt concentrations beyond a thin channel regime the ratio $K_{0}/K^{\infty}$ slowly decays to unity. The low salt plateau is not observed if $\phi_s = 5$ since, as discussed above, the thin film regime is not reached. However, even assuming it could, $K_{0}/K^{\infty}$ would be below  $\cosh (5) \simeq 74$, i.e. this channel cannot provide even two order of magnitude conductivity enhancement. We also remark, that the point of $\ell_{Du}/H = 1$ approximately corresponds to an order of magnitude enhancement of $K_0$ in Fig.~\ref{fig:conduction_CP}, by confirming that when $\ell_{Du}$ exceeds the channel thickness the surface conductivity dominates over the bulk.

The scaling of $K_0$ with salt, especially at low concentrations, is of interest. In the CC channel $K_0 \propto c_{\infty}^{0}$ (see Fig.~\ref{fig:conductivity} and its description in the text). As clarified in Sec.~\ref{sec:CCCP}, in the CP case and a thick channel regime $\ell_{Du}\propto c_{\infty}^{-1/2}$, but for a thin channel regime $\ell_{Du}$ does not depend on salt. Correspondingly, in extremely dilute solutions $K_0 \propto \sqrt{c_{\infty}}$ and $K_0 \propto c_{\infty}$ as follows from Eqs.\eqref{eq:hydrophilic_thick3} and \eqref{eq:hydrophilic_thin2}. This implies that any charge regulation model should inevitably lead to a power-law scaling of $K_0$ (with salt concentration) with an exponent confined between 0 and 1 if the thin channel regime is reached, or between 0 and 1/2 in the thick channel regime. We note that both a 1/2 power-law derived by \citet{biesheuvel.pm:2016} and a 1/3 power-law scaling published by \citet{secchi.e:2016} fall in between these attainable bounds, i.e. both exponents are within a permitted range and not forbidden.

\section{Hydrophobic channels}\label{sec:Hydrophobic_results}

In the case of hydrophobic channels the conductivity can be obtained from Eq.\eqref{eq:M22}, i.e. by summing up $K_0$ and $\Delta K$. More precisely, $\Delta K$ defined by Eq.\eqref{eq:slip correction} should be added to Eq.\eqref{eq:K0} in the thick channel regime, or to  \eqref{eq:hydrophilic_thin} in the thin channel regime.

A limiting case of special interest is that of $\mu = 1$. It follows from \eqref{eq:slip correction} that if all adsorbed charges are immobile, only a convection contribution is superimposed with $K_0$
\begin{equation}\label{eq:slip correction_conv}
  \Delta K = K^{\infty }\frac{12 \ell_{Du}}{H} \frac{ b}{\ell
_{GC}}\dfrac{\mathcal{R}}{\ell _{B}}
\end{equation}
Eq.\eqref{eq:slip correction_conv} indicates that the this contribution is proportional to $\ell_{Du}/{H}$ and scales with $b/\ell
_{GC}$. This suggests that $K$ can be significantly amplified by hydrodynamic slippage provided $\ell_{Du}/{H}$ is not too small and surfaces are highly charged.

\begin{figure}[h]
\begin{center}
\includegraphics[width=1\columnwidth]{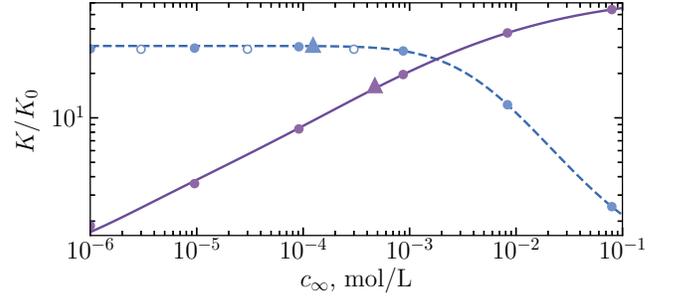}
\end{center}
\caption{Conductivity amplification due to hydrophobicity, $K/K_{0}$, calculated for the channel of $H = 100$ nm using $b = 100$ nm and $\mu = 1$, and fixed $\phi_{s} = 4$ (solid curve) or fixed $\ell_{GC} = 2$ nm (dashed curve). Filled circles are obtained using Eqs.~\eqref{eq:slip correction} and \eqref{eq:hydrophilic_thick}. Open circles are calculations from Eq.~\eqref{eq:K/K0}. The big triangles mark the points of $\ell_{Du}/{H} = 1$.}
\label{fig:slip}
\end{figure}

This case is illustrated in Fig.~\ref{fig:slip}, where  $K/K_0 = 1 + \Delta K/K_0$ for the CC and CP channels is plotted as a function of $c_{\infty}$. The calculations
are made using $\ell_{GC} = 2$ nm and $\phi_s = 4$ ($\Phi_s \simeq 100$ mV). These values provide the regime of the thick channel in the chosen range of $c_{\infty}$. We set $b = 100$ nm that provides constant $b/\ell_{GC} = 50$ in the CC case. In the CP case the ratio $b/\ell_{GC}$ increases with $c_{\infty}$ (see Table~\ref{table:CPCC}) and with our value of $\phi_s$  can be approximated as $b/\ell_{GC} \simeq 3.62 b/\lambda_D$.

For the CC channel $\Delta K$ does not depend on salt and $K/K_0$ takes its
maximal (constant, i.e. independent on $c_{\infty}$) values when $\ell_{Du}/{H} \geq 1$. Dividing Eq.\eqref{eq:slip correction_conv} by Eq.\eqref{eq:hydrophilic_thick3} we obtain for the plateau branch
\begin{equation}\label{eq:K/K0}
\dfrac{K}{K_0} \simeq 1 + \dfrac{b}{\ell _{GC}} \times \dfrac{3 \mathcal{R}}{\ell _{B} + 3 \mathcal{R}},
\end{equation}
which gives ca. 28 for our parameters. Note that $K/K_0$ given by Eq.\eqref{eq:K/K0} does not depend on the channel thickness $H$. The theoretical calculations provide a good estimate of the $K/K_0$ plateau magnitude, and the enhancement compared to a hydrophilic channel is very large, a few tens of times! Using scaling arguments \citet{bocquet.l:2010} concluded that when a saturation plateau of $K_0$ occurs, $\Delta K/ K_0 \propto b/\ell
_{GC}$. Thus, our theoretical, \eqref{eq:K/K0}, and numerical results generally confirm  this scaling.
At larger concentrations, where $\ell_{Du}/{H} < 1$, $K/K_0$ decreases with salt. When  $K_0 \simeq K^{\infty }$, the ratio of Eqs.\eqref{eq:slip correction_conv} and \eqref{eq:ic0_thick} gives $K/K_0 \propto c_{\infty}^{-1}$. We remark and stress that in the intermediate (between the plateau and bulk) region the value of $K/K_0$ is quite large. Therefore, a hydrodynamic slip can provide a significant conductivity enhancement even when $\ell_{Du}/{H}$ is quite small.

By contrast, for a CP channel the value of $K/K_0$ increases with salt as seen in Fig.~\ref{fig:slip}. That it should be so is immediately evident if we recall that  $\ell_{GC} \propto \lambda_D$ and $\ell_{Du}/\ell_{GC}$ is constant (see Sec.~\ref{sec:CCCP} and Table~\ref{table:CPCC}). The latter implies that $\Delta K$ defined by Eq.\eqref{eq:slip correction_conv} increases linearly with $c_{\infty}$ (since linear in $K_{\infty}$). At low salt $\ell_{Du}/H$ is large and Eq.\eqref{eq:K/K0} is valid. It follows that
$K/K_0 \propto \dfrac{b}{\lambda_D} \sinh \dfrac{\phi_s}{2} \propto c_{\infty}^{1/2}$. Since $\lambda_D$ is large in dilute solutions the conductivity enhancement compared to $K_0$ is moderate, even for large slip length of 100 nm taken here. At high salt, where $\ell_{Du}/H$ is small and $K_0 \simeq K_{\infty}$ should saturate to
\begin{equation}\label{eq:CP_sat}
 \dfrac{K}{K_0} \simeq  1 + \dfrac{12 b}{H}  \dfrac{\mathcal{R}}{\ell_{B}}  \sinh^2\dfrac{\phi_s}{2},
\end{equation}
which does not depend on $c_{\infty}$ and is set by $\phi_s$ and $b/H$, so it can be very large for slippery nanochannels of a high surface potential. With our parameters the saturation is not reached, but yet the trend is well seen in Fig.~\ref{fig:slip}. A key remark is that the
amplification of $K$ compared to a bulk conductivity $K_{\infty}$ is solely due to a hydrodynamic slip, but not due to enhanced conductivity in the EDL regions, which are too thin in this case to affect $K$.

\begin{figure}[h]
\begin{center}
\includegraphics[width=1\columnwidth]{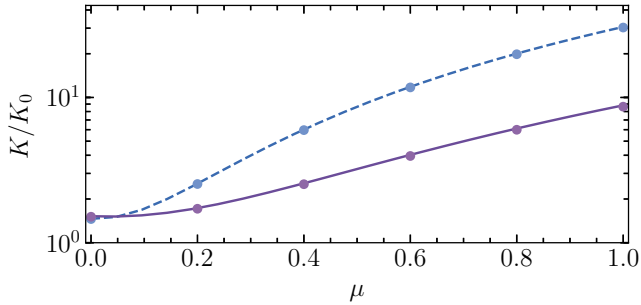}
\end{center}
\caption{$K/K_{0}$  computed at fixed $c_{\infty} = 10^{-4}$ mol/L for the same hydrophobic CC (dashed curve) and CP (solid curve) channels as in Fig.~\ref{fig:slip}, but for a case where only a  fraction $\mu$ of  adsorbed at the walls charges is immobile. The circles are calculations from Eqs.\eqref{eq:slip correction} and \eqref{eq:hydrophilic_thick}.  }
\label{fig:slip_mu}
\end{figure}

We now illustrate the influence of a hydrodynamic slip on conductivity in the case of $\mu \neq 1$. It follows from Eq.\eqref{eq:slip correction} that when all adsorbed charges are mobile, only a migration slip-driven contribution at the walls remains
\begin{equation}\label{eq:slip correction_migr}
 \Delta K = K^{\infty }\frac{4\ell_{Du}}{H}
\end{equation}
Note that although it takes its origin in a hydrodynamic slip, the slip length $b$ does not appear as a parameter in Eq.\eqref{eq:slip correction_migr}. In this case of $\mu = 0$ it follows from Eqs.\eqref{eq:hydrophilic_thick3} and \eqref{eq:slip correction_migr} that for the branch of a (CC) conductivity plateau $\Delta K/K_0 \simeq  \ell_{B}/( \ell_{B} + 3 \mathcal{R})$. With the same parameters as in Fig.~\ref{fig:slip} this gives $K/K_0 \simeq 1.44$, i.e.  when all absorbed ions are mobile, the amplification becomes very  small and can be neglected. If we keep $c_{\infty} = 10^{-4}$ mol/L fixed (at this concentration $\ell_{Du}$ exceeds $H$ as seen in Fig.~\ref{fig:slip}) but increase $\mu$ (decrease the fraction of mobile surface charges), we obtain a situation shown in Fig.~\ref{fig:slip_mu}. On increasing $\mu$, $K/K_0$ computed for a CC channel (of $b/\ell _{GC} = 50$) has a weakly pronounced minimum at $\mu \simeq 0.01$, which is practically not discernible in the scale of Fig.~\ref{fig:slip_mu}, and then quickly increases taking the largest value at $\mu = 1$. In this situation, i.e. when all adsorbed charges are immobile, $K$ exceeds $K_0$ in several tens of times. For the CP channel the trend is the same, but the minimum is slightly shifted (to $\mu \simeq 0.03$) and the enhancement is  much smaller. As a result, one order of magnitude amplification of $K$ is reached only when $\mu = 1$. That the CP channel with $\phi_s = 4$ at given $c_{\infty} = 10^{-4}$ mol/L should provide a smaller $K/K_0$ than the CC channel of $\ell_{B} = 2$ nm is simply a consequence of the reduction of $b/\ell _{GC}$ to ca 12 at a given $c_{\infty}$.  Also included in Fig.~\ref{fig:slip_mu} are theoretical calculations, which are well consistent with the numerical data.

\begin{figure}[h]
\begin{center}
\includegraphics[width=1\columnwidth]{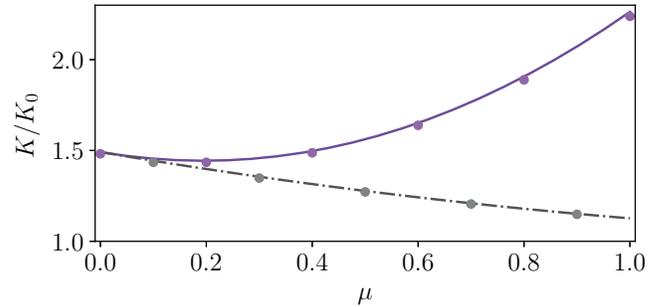}
\end{center}
\caption{The same as in Fig.~\ref{fig:slip_mu}, but computed for a case where $\ell_{GC} = 5$ nm is larger and $b = 10$ (solid curve) and 1 nm (dash-dotted curve) is much smaller so that   $K/K_0$ exhibits a minimum shifted towards a larger $\mu$ or has no minimum. The circles are calculated from Eqs.\eqref{eq:slip correction} and \eqref{eq:hydrophilic_thick}.  }
\label{fig:slip_mu_regimes}
\end{figure}

The results appropriate to Fig.~\ref{fig:slip_mu} refer to large $b/\ell _{GC}$. If we keep the same $c_{\infty} = 10^{-4}$ mol/L fixed, but reduce $b/\ell _{GC}$, the picture can become different, as it is seen in Fig.~\ref{fig:slip_mu_regimes}. Here we limit ourselves by the CC case only, and make numerical calculations using $\ell_{GC} = 5$ nm and two values of $b$, namely, 10 and 1 nm. For an upper curve of  $b/\ell _{GC} = 2$ the minimum is shifted towards larger $\mu$ compared to an example displayed in Fig.~\ref{fig:slip_mu} and occurs at $\mu \simeq 0.19$. The value of $K/K_0$ remains largest at $\mu = 1$, but is quite small by exceeding the value at $\mu = 0$ only in ca.1.5 times. When $b/\ell _{GC} = 0.2$ (a lower curve) no minimum of $K/K_0$ exists  in the range of $\mu$ from 0 to 1. In this circumstances, $K/K_0$  monotonously decreases with $\mu$. That it should be so follows from general considerations. When $\mu = 0$, the fluid velocity at the wall associated with diffuse ions is fully compensated by that associated with mobile adsorbed ions. As a result, a slip-driven convective contribution to the conductivity disappears, but the migration contribution is maximized and is given by Eq.\eqref{eq:slip correction_migr}. On increasing $\mu$ the migration term in the brackets of Eq.\eqref{eq:slip correction} reduces from 1 down to 0, but the convective contribution increases from 0 to its largest value. It follows from Eq.\eqref{eq:slip correction} that for $b/\ell _{GC} < \ell _{B}/3 \mathcal{R} \simeq 0.78$ an extra convective conductivity due to slippage at $\mu = 1$ is even smaller than the migration contribution at $\mu = 0$.
 These results provide new physical insight into an impact of mobile adsorbed ions into the channel conductivity, but certainly at low $b/\ell _{GC}$ a significant enhancement of conductivity is simply not possible. It becomes now evident that a dramatic conductivity enhancement due to slippage can be achieved only provided $b/\ell
_{GC} \gg 1$.

It follows from above results that the largest conductivity amplification compared to $K^{\infty }$ should be expected in a thick channel regime when $\ell_{Du}/\ell_{GC} \gg 1$. In this situation \eqref{eq:hydrophilic_thick2} becomes accurate for a hydrophilic channel. One can then derive a compact equation for a total conductivity of an electrolyte solution confined in a hydrophobic channel

\begin{equation}\label{eq:slip_thick}
K \simeq  K^{\infty } \left[ 1 + \frac{4 \ell_{Du}}{H} \left(\frac{3 \mathcal{R}}{\ell_{B}} \left(1 + \frac{ \mu ^{2}b}{\ell
_{GC}}\right) + 2 - \mu)\right)\right]
\end{equation}

If the surface conductivity dominates significantly over the bulk one, the last equation reduces to
\begin{equation}\label{eq:sc}
K \simeq  K^{\infty } \dfrac{4 \ell_{Du}}{ H} \left[ \frac{3 \mathcal{R}}{\ell_{B}} \left(1 + \frac{ \mu ^{2}b}{\ell
_{GC}}\right) + 2 - \mu \right],
\end{equation}
which is easy to use.
The form of \eqref{eq:sc} suggests that the surface conductivity is linear in $\ell_{Du}/H$, and is also defined by $b/\ell
_{GC} $ and $\mu$.
The contribution of the surface conductivity also grows with $3  \mathcal{R}/\ell_{B}$, which is not small (ca. 1.25 with our parameters), but this ratio characterizes an electrolyte solution and does not depend on the surface properties.

The examples described so far correspond to the thick channel regime, but using Eqs.\eqref{eq:M22}, \eqref{eq:slip correction}, and \eqref{eq:hydrophilic_thin2} it is easy to find that the conductivity in the thin channel  regime at $\ell_{Du}/H \geq 1$ is also given by Eq.\eqref{eq:sc}. Then for CC and CP channels  we obtain
\begin{equation}\label{eq:thin_enhancement}
  \dfrac{K}{K_0} \simeq 2 + \dfrac{3 \mathcal{R}}{\ell _{B}} \dfrac{b \mu^2}{\lambda_D} \dfrac{\lambda_D}{\ell _{GC}}  - \mu \simeq  2 + \dfrac{3 \mathcal{R}}{4 \ell _{B}} \dfrac{b H \mu^2}{\lambda_D^2} \sinh \phi_s  - \mu
\end{equation}
Since this regime implies that  $\lambda_D/\ell _{GC} \leq 1$ and $H \ll \lambda_D$, we conclude that large $K/K_0$ is possible only if $b/\lambda_D \gg 1$. This condition is unlikely to be realised in dilute solutions. To give an idea on possible conductivity enhancement, for the plateau branch of the lower curve (of $\ell_{GC} = 300$ nm) in Fig.~\ref{fig:conductivity}, assuming
$b = 100$ nm  we evaluate  $K/K_0 \simeq 1.2$ at $\mu = 1$ and $K/K_0 \simeq 2$ when $\mu = 0$. Thus, the hydrodynamic slip does not practically affect the conductivity in the thin channel regime.

Finally we return to ``experimental'' $J-E$ curves shown in Fig.~\ref{fig:conduction_current}. The ``hydrophilic'' CC curves are well fitted by Eqs.\eqref{eq:hydrophilic_thick} and \eqref{eq:hydrophilic_thick3} confirming the accuracy of our theory. The calculations from Eq.\eqref{eq:sc} are compared with numerical results for a hydrophobic CP channel. It is seen that they provides a reasonable fit to the data although here $\ell_{Du}/H$ is smaller than 1. We thus conclude that Eq.\eqref{eq:sc} could be a sensible approximation even outside of the range of its formal applicability.

\section{Concluding remarks}\label{sec:conclusion}

We have presented a general theory that incorporates a hydrodynamic slip and a mobility of adsorbed
charges to describe the superenhanced conductivity of an electrolyte solution confined in nanochannels. Numerical solutions are presented and fully validate our analysis. These results are relevant for recent experiments on a
conductivity in nanoslits and nanotubes, which currently become the area of very active research. Our theory provides a direct physical explanation of experimental and numerical results, and also makes specific predictions for a conductivity amplification caused by the variation of various surface properties.

Our results suggest that the electrostatic disjoining pressure $\Pi$ is an important quantity, which determines two different regimes  for a conductivity of a confined electrolyte solution. When $\Pi$ is much smaller than an excess osmotic pressure at the walls, which  often happens in the case of highly charged surfaces even if EDLs strongly overlap, the channel effectively behaves as thick. When $\Pi$ is comparable with  an excess osmotic pressure at the walls, which is possible only when EDLs strongly overlap if and only if surfaces are relatively weakly charged, another limiting case that we referred to as a thin channel regime occurs.
Our model leads to a number of  asymptotic approximations, which are both simple and very accurate, and
provides considerable insight into conductivities in the thick and thin channel regimes.

In the thick channel regime and hydrophilic walls we predict a significant deviations from the bulk conductivity when $H \le \ell_{Du}$, where an electrostatic length $\ell_{Du}$ is defined by Eq.\eqref{eq:dukhin_length}. Since $\ell_{Du}$ can be much larger than any conceivable Debye length, the large surface conductivity does not implies that EDLs necessarily overlap. That a conductance plateau can be obtained without EDL overlap has been first reported by \citet{schoch.rb:2005}. Our theory confirms this experimental result and points out that for ``no-slip'' CC channels the plateau emerges when $\ell_{Du}/H \geq 1$. In the case of CP channels such values of $\ell_{Du}/H$ normally indicate that there is at least an order of magnitude conductivity enhancement, but the later always limited by $\cosh \phi_s$. For the thin channel regime with no-slip at the walls we argue that the conductivity is amplified in $\Pi$ times provided $H$ is the smallest length in the system. Consequently, a large conductivity enhancement is possible for weakly charged CC channels, but not in the CP situation, where $\Pi$ tends to a constant value as $H \to 0$. In the latter case the surface conductivity exceeds that in the bulk, but remains very low. The results obtained for CC and CP cases represent as rigorous bounds on the
attainable conductivity of any hydrophilic channels. This allowed us to specify a possible range  of exponents of a power-law scaling of conductivity with $c_{\infty}$, which remains a subject of hot debates~\cite{secchi.e:2016,biesheuvel.pm:2016}.

The channel conductivity can be further amplified by interfacial slippage, and we have presented a theory of a slip-driven contribution. The latter incorporate a constant hydrodynamic slip length of the surfaces and a finite mobility of adsorbed ions, and is valid for channels of any thickness. Our work clarify the nature of the conductivity enhancement and makes connection with the early studies~\cite{catalano.j:2016,bocquet.l:2010,mouterde.t:2018}. We show that a massive amplification of the conductivity due to hydrodynamic slippage is possible, but only in the thick channel regime and large $b/\ell _{GC}$. In this situation the ratio of conductivity of a slippery channel to that of an equivalent hydrophilic channel takes its minimum value at some (very small) fraction of immobile adsorbed charges $\mu$ and then quickly (nonlinearly) increases with $\mu$. The largest, a few tens of times, conductivity enhancement over the ``no-slip'' case then occurs when all adsorbed charges are immobile. Our work  clarifies that the nature of such a non-monotonous dependence of the slip-driven contribution to a conductivity on the fraction of immobile adsorbed charges reflects a competition between  migration and  convective  terms in the superimposed slip correction given by Eq.\eqref{eq:slip correction}. However, for small $b/\ell _{GC}$ or for a thin channel regime we do not expect a  discernible conductivity enhancement due to slip since an extra convective conductivity is dramatically suppressed.

Our results can be immediately applied for parallel-plate nanochannels. Cylinders should constitute a more realistic model for artificial nanotubes and real porous materials and our calculations are in progress for these. Preliminary results suggest some important quantitative difference between cylinders and slits, but the qualitative features of the conductivity curves are the same.

The dependence of conductivity on static and dynamic surface properties, such as surface charge and potential, fraction of mobile surface charges, and slip length, opens new strategies to tune the ion transport in nanochannels via a modifications of their walls. This has been already used by \citet{karnik.r:2005,karnik.r:2005b} to develop a nanofluidic transistor and to affect the conductance of nanochannels by their surface modification. Our results show that due to a rich physics of the phenomenon it could be many other possibilities to employ the surface properties to control the conductivity. Clearly, more experiments and theoretical analysis is required. It would be of some interest to explore the conductivity in the conical hydrophobic channels to gain insights into the novel physical mechanism controlling current rectification in a geometric diode~\cite{cervera.j:2006,dalcengio.s:2019}.
Our model can also be extended to calculate and optimise the streaming conductivity in hydrophobic nanofluidic channels, which is important for improvement of the energy conversion~\cite{ren.yq:2008}. Another fruitful direction would be diffusio-osmotic energy conversion~\cite{siria.a:2013}, which remains poorly understood
even for hydrophilic channels. Some researchers see this as a promising renewable energy source~\cite{zhang.z:2021}. Other authors, however, argue that salinity gradient generated in full-scale nanoporous membranes is not viable for power generation~\cite{wang.l:2021}. To shed more light on the prospect of harvesting salinity gradient  energy, more theoretical studies are necessary, and it would be timely to include the combined effect of slippage and surface charge mobility into consideration.

\begin{acknowledgments}

 This work was supported by the Ministry of Science and Higher Education of the Russian Federation.
\end{acknowledgments}

\appendix

\section{Derivation of equations for $\overline{(\phi^{\prime} )^{2}}$ and $\overline{\cosh \phi}$}\label{a:2}

In this Appendix we calculate the mean square derivative of the electrostatic potential, and  the mean osmotic pressure in the thick an thin channel regimes. It follows from \eqref{eq:PB_out1} that they are always related as
\begin{equation}\label{eq:rel_phi_cosh}
  \overline{(\phi^{\prime})^2} = \dfrac{2}{ \lambda_D^2} \left[\overline{\cosh \phi}- \Pi - 1 \right]
\end{equation}

\subsection{Thick channel regime}

Substituting \eqref{eq:PB_out2} into \eqref{eq:average} we obtain

\begin{equation}
  \overline{(\phi^{\prime})^{2}} = \dfrac{8}{H \lambda_D^2}\int_{0}^{H/2} \sinh^2(\dfrac{\phi}{2}) dz
\end{equation}
Then making a change of variables and performing the integration we find
\begin{equation}
  \overline{(\phi^{\prime} )^{2}} =  \dfrac{4}{H \lambda_D}\int_{0}^{\phi_s} \sinh(\dfrac{\phi}{2}) d \phi = \dfrac{8}{H \lambda_D} (\cosh\dfrac{\phi}{2} - 1)
\end{equation}
Making use of \eqref{eq:grahame_Du} the last equation can be transformed to Eq.\eqref{eq:average_dphi}.

It follows from \eqref{eq:PB_out3} and \eqref{eq:PB_out2} that
\begin{equation}
  \cosh \phi = 1 + \dfrac{\lambda_D^2 ( \phi^{\prime})^2}{2} = 1 + 2\sinh^2(\dfrac{\phi}{2}),
\end{equation}
and after a change of variables it is straightforward to show that
\begin{equation}
 \overline{\cosh\phi} =  1 + \dfrac{2 \lambda_D}{H }\int_{0}^{\phi_s} \sinh(\dfrac{\phi}{2}) d \phi
\end{equation}
Performing the integration and substituting \eqref{eq:grahame_Du} we obtain Eq.\eqref{eq:average_cosh}. It is easy to verify, that Eqs.\eqref{eq:average_dphi} and \eqref{eq:average_cosh} satisfy \eqref{eq:rel_phi_cosh} when $\Pi$ can be neglected.

\subsection{Thin channel regime}

It follows from \eqref{eq:series} that in a thin channel

\begin{equation}\label{eq:der_thin}
(\phi^{\prime})^2 = \dfrac{z^2 \sinh^2 \phi _{s}}{\lambda _{D}^{4}}
\end{equation}
Therefore,
\begin{equation}
  \overline{(\phi^{\prime} )^{2}} = \dfrac{2\sinh^2 \phi _{s}}{H \lambda _{D}^{4}}\int_{0}^{H/2} z^2 dz = \dfrac{H^2 \sinh^2 \phi _{s}}{12 \lambda _{D}^{4}}
\end{equation}
Using Eq.\eqref{eq:grahame_thin} one can then obtain Eq.\eqref{eq:series_der_thin}.

Making use of the integration by parts formula we can express $\overline{\cosh\phi}$ as
\begin{equation}\label{eq:average_cosh_thin0}
\overline{\cosh\phi} =  \cosh\phi_s - \frac{2}{H}\int\limits_{0}^{H/2} z (\cosh\phi)^{\prime} dz,
\end{equation}

Substituting \eqref{eq:der_thin} into Eq.\eqref{eq:PB_out1} one can obtain

\begin{equation}
\cosh \phi  \simeq \cosh \phi_m + \frac{\sinh^2 \phi_s}{2 \lambda_D^2} z^2,
		\end{equation}%
whence
\begin{equation}
 (\cosh \phi)^{\prime} \simeq  \frac{\sinh^2 \phi_s}{ \lambda_D^2} z,
		\end{equation}%
which can be substituted into \eqref{eq:average_cosh_thin0}. Integrating and using Eq.\eqref{eq:grahame_thin} one obtains Eq.\eqref{eq:average_cosh_thin}.

Substituting Eqs.\eqref{eq:series_der_thin} and \eqref{eq:average_cosh_thin} into \eqref{eq:rel_phi_cosh} we verified the correctness of our calculations.

\section{Derivation of general expressions for conductivity}\label{a:1}

\begin{figure}
\begin{center}
\includegraphics[width=0.99\columnwidth , trim=0.cm 0. 0.0cm
0.,clip=false]{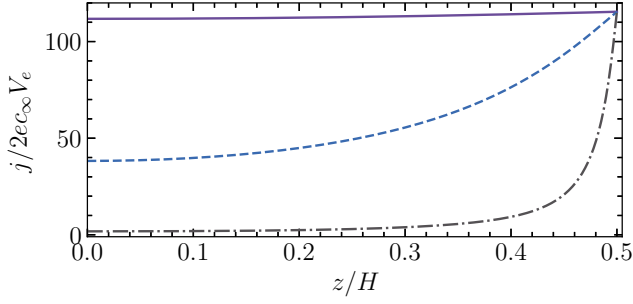}
\end{center}
\caption{Local current density calculated numerically for the same $\phi_s$ and $H/\lambda_{D}$ as in Fig.~\ref{fig:potential_profile} using $\mu = 1$ and $b = 0$. }
\label{fig:j}
\end{figure}

The local current density $j$ divided by $2 e c_{\infty} V_e$, where $V_e = \dfrac{e}{4\pi \eta
\ell _{B}}$, computed with no-slip at the walls for the same nanochannels as in Fig.~\ref{fig:potential_profile} is shown in Fig.~\ref{fig:j}. It can be seen that $j$ is non-uniform and reflects the form of $\phi$. In this Appendix we calculate $J$ in order to derive general expressions for $K_0$ and $\Delta K$ that are valid for any electrostatic channel thickness $H/\lambda_D$.

The local current densities of diffuse ions in Eq.~\eqref{eq:J} can be reexpressed as
%\begin{widetext}
\begin{eqnarray}
\overline{j_{+}} &=& \dfrac{2e V_{e} E c_{\infty}}{H} \int_{0}^{H/2} e^{-\phi} \left[ v + \frac{m_{i}}{V_{e}}  \right] dz \label{eq:j+} \\
\overline{j_{-}} &=& -\dfrac{2eV_{e} E c_{\infty}}{H} \int_{0}^{H/2} e^{\phi} \left[v - \frac{m_{i}}{V_{e}}  \right] dz \label{eq:j-},
\end{eqnarray}
%\end{widetext}
 where $v$ is given by Eq.\eqref{eq:Stokes_solution}.

The convective contribution, $\overline{j_{+}}+\overline{j_{-}}$, to the current density in a hydrophilic channel is then
\begin{equation}\label{eq:conv_j}
  \dfrac{4e V_{e} E c_{\infty}}{H} \left[ \phi_{s} \int_{0}^{H/2}  \sinh\phi dz - \int_{0}^{H/2} \phi \sinh\phi  dz   \right] =  \frac{eV_{e}E}{4\pi\ell_{B}} \overline{ (\phi^{\prime})^{2}}
\end{equation}
Here we used that $\sinh\phi$ is related to $\phi^{\prime \prime}$ by the Poisson-Bolzmann equation \eqref{eq:NLPB}, performing then integration by part.

The migration term for a ``no-slip'' channel reads
	\begin{equation}
	\dfrac{- 2e V_{e} E c_{\infty}m_{i}E}{H}  \int_{0}^{H/2}  \left( e^{-\phi}  + e^{\phi}  \right) dz = \frac{eV_{e}E}{6\pi\lambda_{D}^{2}\mathcal{R}} \overline{\cosh\phi}
	  \label{eq:j_4}
	\end{equation}

Summing up the equations \eqref{eq:conv_j} and \eqref{eq:j_4}, we obtain the conductivity of a hydrophobic channel
\begin{equation}\label{eq:sum1}
 K_0 =  \frac{eV_{e}E}{4\pi\ell_{B}} \overline{ (\phi^{\prime})^{2}} + \dfrac{eV_{e}E}{6\pi\lambda_{D}^{2}\mathcal{R}} \overline{\cosh\phi}
\end{equation}
For uncharged walls the electro-osmotic flow is not generated ($\phi_s = 0$, $\overline{ (\phi^{\prime})^{2}} = 0$, and $\overline{\cosh {\phi }} = 1$).   This immediately gives us the conductivity of the bulk electrolyte solution, Eq.\eqref{eq:ic0_thick},
which is proportional to the salt concentration. Eq.\eqref{eq:sum1} can then be transformed to Eq.\eqref{eq:K0}.

The value of $K_0$ depends on $c_{\infty}$, but is not specific to a wetting situation. For a slippery surface an additional contribution $\Delta K$ to conductivity is generally expected.

If $\mu b \neq 0$, the last term in Eq.\eqref{eq:Stokes_solution} should inevitably lead to an extra convective conductivity. Performing calculations similar to above we find

	\begin{equation}
	\dfrac{-4e V_{e}E c_{\infty}\mu b}{H \ell_{GC}} \int_{0}^{H/2}  \left( e^{-\phi}  - e^{\phi}  \right) dz =  \dfrac{e V_{e} E }{\pi\ell_{B} H} \frac{ 2 \mu b}{\ell_{GC}^2}   \label{eq:j_3}
	\end{equation}

Finally, the adsorbed cations located at $z=0$ react to the electric field by producing an extra conductivity. For a single wall it can be expressed as
\begin{equation}
 j_{\sigma} = \dfrac{(1-\mu )e }{2 \pi \ell _{GC}\ell _{B}} \left( V|_{z=H/2} + \frac{e E}{6 \pi \eta \mathcal{R}} \right)
\end{equation}

Then the last term in Eq.\eqref{eq:J} is then given by

\begin{equation}
\dfrac{2 j_{\sigma}}{H} =  \dfrac{(1-\mu )e }{H \pi \ell _{GC}\ell _{B}} \left( - \dfrac{e E}{4\pi \eta
\ell _{B}} \dfrac{2 \mu b}{\ell _{GC}} + \frac{e E}{6 \pi \eta \mathcal{R}}\right)
\end{equation}

The last equation can be transformed to

	\begin{equation}
\dfrac{2 j_{\sigma}}{H} =  - \frac{eV_{e}E}{\pi\ell_{B}H}\frac{2\mu b}{\ell_{GC}^{2}} + \frac{eV_{e}E}{\pi\ell_{B}H}\frac{2\mu^{2}b}{\ell_{GC}^{2}} + \frac{2eV_{e}E(1-\mu)}{3\pi H \ell_{GC}\mathcal{R}}, \label{eq:j_s}
	\end{equation}

and summing up Eqs.\eqref{eq:j_3} and \eqref{eq:j_s} we find

\begin{equation}
  \Delta K = \frac{2 eV_{e}}{3 \mathcal{R} \pi H \ell_{GC}} \left[\frac{3 \mathcal{R} \mu^{2}b}{\ell_{B}\ell_{GC}} + 1-\mu\right],
\end{equation}
which leads to Eq.\eqref{eq:slip correction}.

\bibliography{current}

\begin{thebibliography}{49}%
\makeatletter
\providecommand \@ifxundefined [1]{%
 \@ifx{#1\undefined}
}%
\providecommand \@ifnum [1]{%
 \ifnum #1\expandafter \@firstoftwo
 \else \expandafter \@secondoftwo
 \fi
}%
\providecommand \@ifx [1]{%
 \ifx #1\expandafter \@firstoftwo
 \else \expandafter \@secondoftwo
 \fi
}%
\providecommand \natexlab [1]{#1}%
\providecommand \enquote  [1]{``#1''}%
\providecommand \bibnamefont  [1]{#1}%
\providecommand \bibfnamefont [1]{#1}%
\providecommand \citenamefont [1]{#1}%
\providecommand \href@noop [0]{\@secondoftwo}%
\providecommand \href [0]{\begingroup \@sanitize@url \@href}%
\providecommand \@href[1]{\@@startlink{#1}\@@href}%
\providecommand \@@href[1]{\endgroup#1\@@endlink}%
\providecommand \@sanitize@url [0]{\catcode `\\12\catcode `\$12\catcode
  `\&12\catcode `\#12\catcode `\^12\catcode `\_12\catcode `\%12\relax}%
\providecommand \@@startlink[1]{}%
\providecommand \@@endlink[0]{}%
\providecommand \url  [0]{\begingroup\@sanitize@url \@url }%
\providecommand \@url [1]{\endgroup\@href {#1}{\urlprefix }}%
\providecommand \urlprefix  [0]{URL }%
\providecommand \Eprint [0]{\href }%
\providecommand \doibase [0]{http://dx.doi.org/}%
\providecommand \selectlanguage [0]{\@gobble}%
\providecommand \bibinfo  [0]{\@secondoftwo}%
\providecommand \bibfield  [0]{\@secondoftwo}%
\providecommand \translation [1]{[#1]}%
\providecommand \BibitemOpen [0]{}%
\providecommand \bibitemStop [0]{}%
\providecommand \bibitemNoStop [0]{.\EOS\space}%
\providecommand \EOS [0]{\spacefactor3000\relax}%
\providecommand \BibitemShut  [1]{\csname bibitem#1\endcsname}%
\let\auto@bib@innerbib\@empty
%</preamble>
\bibitem [{\citenamefont {Delgado}\ \emph {et~al.}(2007)\citenamefont
  {Delgado}, \citenamefont {Gonz{\'a}lez-Caballero}, \citenamefont {Hunter},
  \citenamefont {Koopal},\ and\ \citenamefont {Lyklema}}]{delgado.av:2007}%
  \BibitemOpen
  \bibfield  {author} {\bibinfo {author} {\bibfnamefont {{\'A}.~V.}\
  \bibnamefont {Delgado}}, \bibinfo {author} {\bibfnamefont {F.}~\bibnamefont
  {Gonz{\'a}lez-Caballero}}, \bibinfo {author} {\bibfnamefont {R.}~\bibnamefont
  {Hunter}}, \bibinfo {author} {\bibfnamefont {L.}~\bibnamefont {Koopal}}, \
  and\ \bibinfo {author} {\bibfnamefont {J.}~\bibnamefont {Lyklema}},\
  }\href@noop {} {\bibfield  {journal} {\bibinfo  {journal} {J. Colloid
  Interface Sci.}\ }\textbf {\bibinfo {volume} {309}},\ \bibinfo {pages} {194}
  (\bibinfo {year} {2007})}\BibitemShut {NoStop}%
\bibitem [{\citenamefont {Lyklema}\ and\ \citenamefont
  {Minor}(1998)}]{lyklema.j:1998}%
  \BibitemOpen
  \bibfield  {author} {\bibinfo {author} {\bibfnamefont {J.}~\bibnamefont
  {Lyklema}}\ and\ \bibinfo {author} {\bibfnamefont {M.}~\bibnamefont
  {Minor}},\ }\href@noop {} {\bibfield  {journal} {\bibinfo  {journal}
  {Colloids and Surfaces A: Physicochemical and Engineering Aspects}\ }\textbf
  {\bibinfo {volume} {140}},\ \bibinfo {pages} {33} (\bibinfo {year}
  {1998})}\BibitemShut {NoStop}%
\bibitem [{\citenamefont {Schoch}\ \emph {et~al.}(2008)\citenamefont {Schoch},
  \citenamefont {Han},\ and\ \citenamefont {Renaud}}]{schoch.rb:2008}%
  \BibitemOpen
  \bibfield  {author} {\bibinfo {author} {\bibfnamefont {R.~B.}\ \bibnamefont
  {Schoch}}, \bibinfo {author} {\bibfnamefont {J.}~\bibnamefont {Han}}, \ and\
  \bibinfo {author} {\bibfnamefont {P.}~\bibnamefont {Renaud}},\ }\href@noop {}
  {\bibfield  {journal} {\bibinfo  {journal} {Rev. Mod. Phys.}\ }\textbf
  {\bibinfo {volume} {80}},\ \bibinfo {pages} {839} (\bibinfo {year}
  {2008})}\BibitemShut {NoStop}%
\bibitem [{\citenamefont {Stein}\ \emph {et~al.}(2004)\citenamefont {Stein},
  \citenamefont {Kruithof},\ and\ \citenamefont {Dekker}}]{stein.d:2004}%
  \BibitemOpen
  \bibfield  {author} {\bibinfo {author} {\bibfnamefont {D.}~\bibnamefont
  {Stein}}, \bibinfo {author} {\bibfnamefont {M.}~\bibnamefont {Kruithof}}, \
  and\ \bibinfo {author} {\bibfnamefont {C.}~\bibnamefont {Dekker}},\
  }\href@noop {} {\bibfield  {journal} {\bibinfo  {journal} {Phys. Rev. Lett.}\
  }\textbf {\bibinfo {volume} {93}},\ \bibinfo {pages} {035901} (\bibinfo
  {year} {2004})}\BibitemShut {NoStop}%
\bibitem [{\citenamefont {Schoch}\ \emph {et~al.}(2005)\citenamefont {Schoch},
  \citenamefont {van Lintel},\ and\ \citenamefont {Renaud}}]{schoch.rb:2005}%
  \BibitemOpen
  \bibfield  {author} {\bibinfo {author} {\bibfnamefont {R.~B.}\ \bibnamefont
  {Schoch}}, \bibinfo {author} {\bibfnamefont {H.}~\bibnamefont {van Lintel}},
  \ and\ \bibinfo {author} {\bibfnamefont {P.}~\bibnamefont {Renaud}},\
  }\href@noop {} {\bibfield  {journal} {\bibinfo  {journal} {Phys. Fluids}\
  }\textbf {\bibinfo {volume} {17}},\ \bibinfo {pages} {100604} (\bibinfo
  {year} {2005})}\BibitemShut {NoStop}%
\bibitem [{\citenamefont {Siria}\ \emph {et~al.}(2013)\citenamefont {Siria},
  \citenamefont {Poncharal}, \citenamefont {Biance}, \citenamefont {Fulcrand},
  \citenamefont {Blase}, \citenamefont {Purcell},\ and\ \citenamefont
  {Bocquet}}]{siria.a:2013}%
  \BibitemOpen
  \bibfield  {author} {\bibinfo {author} {\bibfnamefont {A.}~\bibnamefont
  {Siria}}, \bibinfo {author} {\bibfnamefont {P.}~\bibnamefont {Poncharal}},
  \bibinfo {author} {\bibfnamefont {A.-L.}\ \bibnamefont {Biance}}, \bibinfo
  {author} {\bibfnamefont {R.}~\bibnamefont {Fulcrand}}, \bibinfo {author}
  {\bibfnamefont {X.}~\bibnamefont {Blase}}, \bibinfo {author} {\bibfnamefont
  {S.~T.}\ \bibnamefont {Purcell}}, \ and\ \bibinfo {author} {\bibfnamefont
  {L.}~\bibnamefont {Bocquet}},\ }\href@noop {} {\bibfield  {journal} {\bibinfo
   {journal} {Nature}\ }\textbf {\bibinfo {volume} {494}},\ \bibinfo {pages}
  {455} (\bibinfo {year} {2013})}\BibitemShut {NoStop}%
\bibitem [{\citenamefont {Balme}\ \emph {et~al.}(2015)\citenamefont {Balme},
  \citenamefont {Picaud}, \citenamefont {Manghi}, \citenamefont {Palmeri},
  \citenamefont {Bechelany}, \citenamefont {Cabello-Aguilar}, \citenamefont
  {Abou-Chaaya}, \citenamefont {Miele}, \citenamefont {Balanzat},\ and\
  \citenamefont {Janot}}]{balme.s:2015}%
  \BibitemOpen
  \bibfield  {author} {\bibinfo {author} {\bibfnamefont {S.}~\bibnamefont
  {Balme}}, \bibinfo {author} {\bibfnamefont {F.}~\bibnamefont {Picaud}},
  \bibinfo {author} {\bibfnamefont {M.}~\bibnamefont {Manghi}}, \bibinfo
  {author} {\bibfnamefont {J.}~\bibnamefont {Palmeri}}, \bibinfo {author}
  {\bibfnamefont {M.}~\bibnamefont {Bechelany}}, \bibinfo {author}
  {\bibfnamefont {S.}~\bibnamefont {Cabello-Aguilar}}, \bibinfo {author}
  {\bibfnamefont {A.}~\bibnamefont {Abou-Chaaya}}, \bibinfo {author}
  {\bibfnamefont {P.}~\bibnamefont {Miele}}, \bibinfo {author} {\bibfnamefont
  {E.}~\bibnamefont {Balanzat}}, \ and\ \bibinfo {author} {\bibfnamefont
  {J.~M.}\ \bibnamefont {Janot}},\ }\href@noop {} {\bibfield  {journal}
  {\bibinfo  {journal} {Sci. Rep.}\ }\textbf {\bibinfo {volume} {5}},\ \bibinfo
  {pages} {10135} (\bibinfo {year} {2015})}\BibitemShut {NoStop}%
\bibitem [{\citenamefont {Bonhomme}\ \emph {et~al.}(2015)\citenamefont
  {Bonhomme}, \citenamefont {Mounier}, \citenamefont {Simon},\ and\
  \citenamefont {Biance}}]{bonhomme.o:2015}%
  \BibitemOpen
  \bibfield  {author} {\bibinfo {author} {\bibfnamefont {O.}~\bibnamefont
  {Bonhomme}}, \bibinfo {author} {\bibfnamefont {A.}~\bibnamefont {Mounier}},
  \bibinfo {author} {\bibfnamefont {G.}~\bibnamefont {Simon}}, \ and\ \bibinfo
  {author} {\bibfnamefont {A.-L.}\ \bibnamefont {Biance}},\ }\href@noop {}
  {\bibfield  {journal} {\bibinfo  {journal} {J. Phys.: Condens. Matter}\
  }\textbf {\bibinfo {volume} {27}},\ \bibinfo {pages} {194118} (\bibinfo
  {year} {2015})}\BibitemShut {NoStop}%
\bibitem [{\citenamefont {Bonhomme}\ \emph {et~al.}(2017)\citenamefont
  {Bonhomme}, \citenamefont {Blanc}, \citenamefont {Joly}, \citenamefont
  {Ybert},\ and\ \citenamefont {Biance}}]{bonhomme.o:2017}%
  \BibitemOpen
  \bibfield  {author} {\bibinfo {author} {\bibfnamefont {O.}~\bibnamefont
  {Bonhomme}}, \bibinfo {author} {\bibfnamefont {B.}~\bibnamefont {Blanc}},
  \bibinfo {author} {\bibfnamefont {L.}~\bibnamefont {Joly}}, \bibinfo {author}
  {\bibfnamefont {C.}~\bibnamefont {Ybert}}, \ and\ \bibinfo {author}
  {\bibfnamefont {A.-L.}\ \bibnamefont {Biance}},\ }\href@noop {} {\bibfield
  {journal} {\bibinfo  {journal} {Adv. Colloid Interface Sci.}\ }\textbf
  {\bibinfo {volume} {247}},\ \bibinfo {pages} {477} (\bibinfo {year}
  {2017})}\BibitemShut {NoStop}%
\bibitem [{\citenamefont {Levine}\ \emph {et~al.}(1975)\citenamefont {Levine},
  \citenamefont {Marriott},\ and\ \citenamefont {Robinson}}]{levine.s:1975}%
  \BibitemOpen
  \bibfield  {author} {\bibinfo {author} {\bibfnamefont {S.}~\bibnamefont
  {Levine}}, \bibinfo {author} {\bibfnamefont {J.~R.}\ \bibnamefont
  {Marriott}}, \ and\ \bibinfo {author} {\bibfnamefont {K.}~\bibnamefont
  {Robinson}},\ }\href@noop {} {\bibfield  {journal} {\bibinfo  {journal}
  {Journal of the Chemical Society, Faraday Transactions 2: Molecular and
  Chemical Physics}\ }\textbf {\bibinfo {volume} {71}},\ \bibinfo {pages} {1}
  (\bibinfo {year} {1975})}\BibitemShut {NoStop}%
\bibitem [{\citenamefont {Ninham}\ and\ \citenamefont
  {Parsegian}(1971)}]{ninham.bw:1971}%
  \BibitemOpen
  \bibfield  {author} {\bibinfo {author} {\bibfnamefont {B.~W.}\ \bibnamefont
  {Ninham}}\ and\ \bibinfo {author} {\bibfnamefont {V.~A.}\ \bibnamefont
  {Parsegian}},\ }\href@noop {} {\bibfield  {journal} {\bibinfo  {journal} {J.
  Theor. Biol.}\ }\textbf {\bibinfo {volume} {31}},\ \bibinfo {pages} {405}
  (\bibinfo {year} {1971})}\BibitemShut {NoStop}%
\bibitem [{\citenamefont {Chan}\ \emph {et~al.}(1975)\citenamefont {Chan},
  \citenamefont {Perram}, \citenamefont {White},\ and\ \citenamefont
  {Healy}}]{chan.d:1975}%
  \BibitemOpen
  \bibfield  {author} {\bibinfo {author} {\bibfnamefont {D.}~\bibnamefont
  {Chan}}, \bibinfo {author} {\bibfnamefont {J.~W.}\ \bibnamefont {Perram}},
  \bibinfo {author} {\bibfnamefont {L.~R.}\ \bibnamefont {White}}, \ and\
  \bibinfo {author} {\bibfnamefont {T.~W.}\ \bibnamefont {Healy}},\ }\href@noop
  {} {\bibfield  {journal} {\bibinfo  {journal} {J. Chem. Soc.{,} Faraday
  Trans. 1}\ }\textbf {\bibinfo {volume} {71}},\ \bibinfo {pages} {1046}
  (\bibinfo {year} {1975})}\BibitemShut {NoStop}%
\bibitem [{\citenamefont {Secchi}\ \emph {et~al.}(2016)\citenamefont {Secchi},
  \citenamefont {Nigu\`es}, \citenamefont {Jubin}, \citenamefont {Siria},\ and\
  \citenamefont {Bocquet}}]{secchi.e:2016}%
  \BibitemOpen
  \bibfield  {author} {\bibinfo {author} {\bibfnamefont {E.}~\bibnamefont
  {Secchi}}, \bibinfo {author} {\bibfnamefont {A.}~\bibnamefont {Nigu\`es}},
  \bibinfo {author} {\bibfnamefont {L.}~\bibnamefont {Jubin}}, \bibinfo
  {author} {\bibfnamefont {A.}~\bibnamefont {Siria}}, \ and\ \bibinfo {author}
  {\bibfnamefont {L.}~\bibnamefont {Bocquet}},\ }\href@noop {} {\bibfield
  {journal} {\bibinfo  {journal} {Phys. Rev. Lett.}\ }\textbf {\bibinfo
  {volume} {116}},\ \bibinfo {pages} {154501} (\bibinfo {year}
  {2016})}\BibitemShut {NoStop}%
\bibitem [{\citenamefont {Biesheuvel}\ and\ \citenamefont
  {Bazant}(2016)}]{biesheuvel.pm:2016}%
  \BibitemOpen
  \bibfield  {author} {\bibinfo {author} {\bibfnamefont {P.}~\bibnamefont
  {Biesheuvel}}\ and\ \bibinfo {author} {\bibfnamefont {M.}~\bibnamefont
  {Bazant}},\ }\href@noop {} {\bibfield  {journal} {\bibinfo  {journal} {Phys.
  Rev. E}\ }\textbf {\bibinfo {volume} {94}},\ \bibinfo {pages} {050601}
  (\bibinfo {year} {2016})}\BibitemShut {NoStop}%
\bibitem [{\citenamefont {Vinogradova}(1999)}]{vinogradova.oi:1999}%
  \BibitemOpen
  \bibfield  {author} {\bibinfo {author} {\bibfnamefont {O.~I.}\ \bibnamefont
  {Vinogradova}},\ }\href@noop {} {\bibfield  {journal} {\bibinfo  {journal}
  {Int. J. Miner. Process.}\ }\textbf {\bibinfo {volume} {56}},\ \bibinfo
  {pages} {31} (\bibinfo {year} {1999})}\BibitemShut {NoStop}%
\bibitem [{\citenamefont {Vinogradova}\ and\ \citenamefont
  {Belyaev}(2011)}]{vinogradova.oi:2011}%
  \BibitemOpen
  \bibfield  {author} {\bibinfo {author} {\bibfnamefont {O.~I.}\ \bibnamefont
  {Vinogradova}}\ and\ \bibinfo {author} {\bibfnamefont {A.~V.}\ \bibnamefont
  {Belyaev}},\ }\href@noop {} {\bibfield  {journal} {\bibinfo  {journal} {J.
  Phys.: Condens. Matter}\ }\textbf {\bibinfo {volume} {23}},\ \bibinfo {pages}
  {184104} (\bibinfo {year} {2011})}\BibitemShut {NoStop}%
\bibitem [{\citenamefont {Cottin-Bizonne}\ \emph {et~al.}(2005)\citenamefont
  {Cottin-Bizonne}, \citenamefont {Cross}, \citenamefont {Steinberger},\ and\
  \citenamefont {Charlaix}}]{charlaix.e:2005}%
  \BibitemOpen
  \bibfield  {author} {\bibinfo {author} {\bibfnamefont {C.}~\bibnamefont
  {Cottin-Bizonne}}, \bibinfo {author} {\bibfnamefont {B.}~\bibnamefont
  {Cross}}, \bibinfo {author} {\bibfnamefont {A.}~\bibnamefont {Steinberger}},
  \ and\ \bibinfo {author} {\bibfnamefont {E.}~\bibnamefont {Charlaix}},\
  }\href@noop {} {\bibfield  {journal} {\bibinfo  {journal} {Phys. Rev. Lett.}\
  }\textbf {\bibinfo {volume} {94}},\ \bibinfo {pages} {056102} (\bibinfo
  {year} {2005})}\BibitemShut {NoStop}%
\bibitem [{\citenamefont {{Vinogradova}}\ and\ \citenamefont
  {{Yakubov}}(2003)}]{vinogradova.oi:2003}%
  \BibitemOpen
  \bibfield  {author} {\bibinfo {author} {\bibfnamefont {O.~I.}\ \bibnamefont
  {{Vinogradova}}}\ and\ \bibinfo {author} {\bibfnamefont {G.~E.}\ \bibnamefont
  {{Yakubov}}},\ }\href@noop {} {\bibfield  {journal} {\bibinfo  {journal}
  {Langmuir}\ }\textbf {\bibinfo {volume} {19}},\ \bibinfo {pages} {1227}
  (\bibinfo {year} {2003})}\BibitemShut {NoStop}%
\bibitem [{\citenamefont {Joly}\ \emph {et~al.}(2006)\citenamefont {Joly},
  \citenamefont {Ybert},\ and\ \citenamefont {Bocquet}}]{joly.l:2006}%
  \BibitemOpen
  \bibfield  {author} {\bibinfo {author} {\bibfnamefont {L.}~\bibnamefont
  {Joly}}, \bibinfo {author} {\bibfnamefont {C.}~\bibnamefont {Ybert}}, \ and\
  \bibinfo {author} {\bibfnamefont {L.}~\bibnamefont {Bocquet}},\ }\href@noop
  {} {\bibfield  {journal} {\bibinfo  {journal} {Phys. Rev. Lett.}\ }\textbf
  {\bibinfo {volume} {96}},\ \bibinfo {pages} {046101} (\bibinfo {year}
  {2006})}\BibitemShut {NoStop}%
\bibitem [{\citenamefont {Vinogradova}\ \emph {et~al.}(2009)\citenamefont
  {Vinogradova}, \citenamefont {Koynov}, \citenamefont {Best},\ and\
  \citenamefont {Feuillebois}}]{vinogradova.oi:2009}%
  \BibitemOpen
  \bibfield  {author} {\bibinfo {author} {\bibfnamefont {O.~I.}\ \bibnamefont
  {Vinogradova}}, \bibinfo {author} {\bibfnamefont {K.}~\bibnamefont {Koynov}},
  \bibinfo {author} {\bibfnamefont {A.}~\bibnamefont {Best}}, \ and\ \bibinfo
  {author} {\bibfnamefont {F.}~\bibnamefont {Feuillebois}},\ }\href@noop {}
  {\bibfield  {journal} {\bibinfo  {journal} {Phys. Rev. Lett.}\ }\textbf
  {\bibinfo {volume} {102}},\ \bibinfo {pages} {118302} (\bibinfo {year}
  {2009})}\BibitemShut {NoStop}%
\bibitem [{\citenamefont {Vinogradova}(1995)}]{vinogradova.oi:1995a}%
  \BibitemOpen
  \bibfield  {author} {\bibinfo {author} {\bibfnamefont {O.~I.}\ \bibnamefont
  {Vinogradova}},\ }\href@noop {} {\bibfield  {journal} {\bibinfo  {journal}
  {Langmuir}\ }\textbf {\bibinfo {volume} {11}},\ \bibinfo {pages} {2213}
  (\bibinfo {year} {1995})}\BibitemShut {NoStop}%
\bibitem [{\citenamefont {Muller}\ \emph {et~al.}(1986)\citenamefont {Muller},
  \citenamefont {Sergeeva}, \citenamefont {Sobolev},\ and\ \citenamefont
  {Churaev}}]{muller.vm:1986}%
  \BibitemOpen
  \bibfield  {author} {\bibinfo {author} {\bibfnamefont {V.~M.}\ \bibnamefont
  {Muller}}, \bibinfo {author} {\bibfnamefont {I.~P.}\ \bibnamefont
  {Sergeeva}}, \bibinfo {author} {\bibfnamefont {V.~D.}\ \bibnamefont
  {Sobolev}}, \ and\ \bibinfo {author} {\bibfnamefont {N.~V.}\ \bibnamefont
  {Churaev}},\ }\href@noop {} {\bibfield  {journal} {\bibinfo  {journal}
  {Colloid J. USSR}\ }\textbf {\bibinfo {volume} {48}},\ \bibinfo {pages} {606}
  (\bibinfo {year} {1986})}\BibitemShut {NoStop}%
\bibitem [{\citenamefont {Joly}\ \emph {et~al.}(2004)\citenamefont {Joly},
  \citenamefont {Ybert}, \citenamefont {Trizac},\ and\ \citenamefont
  {Bocquet}}]{joly2004}%
  \BibitemOpen
  \bibfield  {author} {\bibinfo {author} {\bibfnamefont {L.}~\bibnamefont
  {Joly}}, \bibinfo {author} {\bibfnamefont {C.}~\bibnamefont {Ybert}},
  \bibinfo {author} {\bibfnamefont {E.}~\bibnamefont {Trizac}}, \ and\ \bibinfo
  {author} {\bibfnamefont {L.}~\bibnamefont {Bocquet}},\ }\href@noop {}
  {\bibfield  {journal} {\bibinfo  {journal} {Phys. Rev. Lett.}\ }\textbf
  {\bibinfo {volume} {93}},\ \bibinfo {pages} {257805} (\bibinfo {year}
  {2004})}\BibitemShut {NoStop}%
\bibitem [{\citenamefont {Silkina}\ \emph {et~al.}(2019)\citenamefont
  {Silkina}, \citenamefont {Asmolov},\ and\ \citenamefont
  {Vinogradova}}]{silkina.ef:2019}%
  \BibitemOpen
  \bibfield  {author} {\bibinfo {author} {\bibfnamefont {E.~F.}\ \bibnamefont
  {Silkina}}, \bibinfo {author} {\bibfnamefont {E.~S.}\ \bibnamefont
  {Asmolov}}, \ and\ \bibinfo {author} {\bibfnamefont {O.~I.}\ \bibnamefont
  {Vinogradova}},\ }\href {\doibase 10.1039/C9CP04259H} {\bibfield  {journal}
  {\bibinfo  {journal} {Physical Chemistry Chemical Physics}\ }\textbf
  {\bibinfo {volume} {21}},\ \bibinfo {pages} {23036} (\bibinfo {year}
  {2019})}\BibitemShut {NoStop}%
\bibitem [{\citenamefont {Maduar}\ \emph {et~al.}(2015)\citenamefont {Maduar},
  \citenamefont {Belyaev}, \citenamefont {Lobaskin},\ and\ \citenamefont
  {Vinogradova}}]{maduar.sr:2015}%
  \BibitemOpen
  \bibfield  {author} {\bibinfo {author} {\bibfnamefont {S.~R.}\ \bibnamefont
  {Maduar}}, \bibinfo {author} {\bibfnamefont {A.~V.}\ \bibnamefont {Belyaev}},
  \bibinfo {author} {\bibfnamefont {V.}~\bibnamefont {Lobaskin}}, \ and\
  \bibinfo {author} {\bibfnamefont {O.~I.}\ \bibnamefont {Vinogradova}},\
  }\href@noop {} {\bibfield  {journal} {\bibinfo  {journal} {Phys. Rev. Lett.}\
  }\textbf {\bibinfo {volume} {114}},\ \bibinfo {pages} {118301} (\bibinfo
  {year} {2015})}\BibitemShut {NoStop}%
\bibitem [{\citenamefont {Grosjean}\ \emph {et~al.}(2019)\citenamefont
  {Grosjean}, \citenamefont {Bocquet},\ and\ \citenamefont
  {Vuilleumier}}]{grosjean.b:2019}%
  \BibitemOpen
  \bibfield  {author} {\bibinfo {author} {\bibfnamefont {B.}~\bibnamefont
  {Grosjean}}, \bibinfo {author} {\bibfnamefont {M.-L.}\ \bibnamefont
  {Bocquet}}, \ and\ \bibinfo {author} {\bibfnamefont {R.}~\bibnamefont
  {Vuilleumier}},\ }\href@noop {} {\bibfield  {journal} {\bibinfo  {journal}
  {Nat. Com.}\ }\textbf {\bibinfo {volume} {10}},\ \bibinfo {pages} {1656}
  (\bibinfo {year} {2019})}\BibitemShut {NoStop}%
\bibitem [{\citenamefont {Catalano}\ \emph {et~al.}(2016)\citenamefont
  {Catalano}, \citenamefont {Lammertink},\ and\ \citenamefont
  {Biesheuvel}}]{catalano.j:2016}%
  \BibitemOpen
  \bibfield  {author} {\bibinfo {author} {\bibfnamefont {J.}~\bibnamefont
  {Catalano}}, \bibinfo {author} {\bibfnamefont {R.~G.~H.}\ \bibnamefont
  {Lammertink}}, \ and\ \bibinfo {author} {\bibfnamefont {P.~M.}\ \bibnamefont
  {Biesheuvel}},\ }\href@noop {} {\bibfield  {journal} {\bibinfo  {journal}
  {arXiv:1603.09293}\ } (\bibinfo {year} {2016})}\BibitemShut {NoStop}%
\bibitem [{\citenamefont {{Bocquet}}\ and\ \citenamefont
  {Charlaix}(2010)}]{bocquet.l:2010}%
  \BibitemOpen
  \bibfield  {author} {\bibinfo {author} {\bibfnamefont {L.}~\bibnamefont
  {{Bocquet}}}\ and\ \bibinfo {author} {\bibfnamefont {E.}~\bibnamefont
  {Charlaix}},\ }\href@noop {} {\bibfield  {journal} {\bibinfo  {journal}
  {Chem. Soc. Rev.}\ }\textbf {\bibinfo {volume} {39}},\ \bibinfo {pages}
  {1073} (\bibinfo {year} {2010})}\BibitemShut {NoStop}%
\bibitem [{\citenamefont {Andelman}(2006)}]{poon.w:2006}%
  \BibitemOpen
  \bibfield  {author} {\bibinfo {author} {\bibfnamefont {D.}~\bibnamefont
  {Andelman}},\ }\enquote {\bibinfo {title} {Soft condensed matter physics in
  molecular and cell biology},}\ \ (\bibinfo  {publisher} {CRC Press},\
  \bibinfo {address} {Boca Raton},\ \bibinfo {year} {2006})\ Chap.\ \bibinfo
  {chapter} {6.~Introduction to Electrostatics in Soft and Biological Matter},\
  \bibinfo {edition} {1st}\ ed.\BibitemShut {Stop}%
\bibitem [{\citenamefont {Mouterde}\ and\ \citenamefont
  {Bocquet}(2018)}]{mouterde.t:2018}%
  \BibitemOpen
  \bibfield  {author} {\bibinfo {author} {\bibfnamefont {T.}~\bibnamefont
  {Mouterde}}\ and\ \bibinfo {author} {\bibfnamefont {L.}~\bibnamefont
  {Bocquet}},\ }\href@noop {} {\bibfield  {journal} {\bibinfo  {journal} {Eur.
  Phys. J. E}\ }\textbf {\bibinfo {volume} {41}},\ \bibinfo {pages} {148}
  (\bibinfo {year} {2018})}\BibitemShut {NoStop}%
\bibitem [{\citenamefont {Israelachvili}(2011)}]{israelachvili.jn:2011}%
  \BibitemOpen
  \bibfield  {author} {\bibinfo {author} {\bibfnamefont {J.~N.}\ \bibnamefont
  {Israelachvili}},\ }\href@noop {} {\emph {\bibinfo {title} {Intermolecular
  and Surface Forces}}},\ \bibinfo {edition} {3rd}\ ed.\ (\bibinfo  {publisher}
  {Academic Press},\ \bibinfo {year} {2011})\BibitemShut {NoStop}%
\bibitem [{\citenamefont {Zhou}\ \emph {et~al.}(2019)\citenamefont {Zhou},
  \citenamefont {Wang}, \citenamefont {Fan}, \citenamefont {Xu},\ and\
  \citenamefont {Yang}}]{zhu.h:2019}%
  \BibitemOpen
  \bibfield  {author} {\bibinfo {author} {\bibfnamefont {H.}~\bibnamefont
  {Zhou}}, \bibinfo {author} {\bibfnamefont {Y.}~\bibnamefont {Wang}}, \bibinfo
  {author} {\bibfnamefont {Y.}~\bibnamefont {Fan}}, \bibinfo {author}
  {\bibfnamefont {J.}~\bibnamefont {Xu}}, \ and\ \bibinfo {author}
  {\bibfnamefont {C.}~\bibnamefont {Yang}},\ }\href@noop {} {\bibfield
  {journal} {\bibinfo  {journal} {Adv. Theory Simul.}\ ,\ \bibinfo {pages}
  {1900016}} (\bibinfo {year} {2019})}\BibitemShut {NoStop}%
\bibitem [{\citenamefont {Bonthuis}\ and\ \citenamefont
  {Netz}(2012)}]{bonthuis.dj:2012}%
  \BibitemOpen
  \bibfield  {author} {\bibinfo {author} {\bibfnamefont {D.~J.}\ \bibnamefont
  {Bonthuis}}\ and\ \bibinfo {author} {\bibfnamefont {R.~R.}\ \bibnamefont
  {Netz}},\ }\href@noop {} {\bibfield  {journal} {\bibinfo  {journal}
  {Langmuir}\ }\textbf {\bibinfo {volume} {28}},\ \bibinfo {pages} {16049}
  (\bibinfo {year} {2012})}\BibitemShut {NoStop}%
\bibitem [{\citenamefont {Ninham}\ \emph {et~al.}(1997)\citenamefont {Ninham},
  \citenamefont {Kurihara},\ and\ \citenamefont
  {Vinogradova}}]{ninham.bw:1997}%
  \BibitemOpen
  \bibfield  {author} {\bibinfo {author} {\bibfnamefont {B.~W.}\ \bibnamefont
  {Ninham}}, \bibinfo {author} {\bibfnamefont {K.}~\bibnamefont {Kurihara}}, \
  and\ \bibinfo {author} {\bibfnamefont {O.~I.}\ \bibnamefont {Vinogradova}},\
  }\href@noop {} {\bibfield  {journal} {\bibinfo  {journal} {Colloid. Surf. A}\
  }\textbf {\bibinfo {volume} {123--124}},\ \bibinfo {pages} {7} (\bibinfo
  {year} {1997})}\BibitemShut {NoStop}%
\bibitem [{\citenamefont {Cao}\ and\ \citenamefont {Netz}(2018)}]{cao.q:2018}%
  \BibitemOpen
  \bibfield  {author} {\bibinfo {author} {\bibfnamefont {Q.}~\bibnamefont
  {Cao}}\ and\ \bibinfo {author} {\bibfnamefont {R.~R.}\ \bibnamefont {Netz}},\
  }\href@noop {} {\bibfield  {journal} {\bibinfo  {journal} {Electrochim.
  Acta}\ }\textbf {\bibinfo {volume} {259}},\ \bibinfo {pages} {1011} (\bibinfo
  {year} {2018})}\BibitemShut {NoStop}%
\bibitem [{\citenamefont {Kavokine}\ \emph {et~al.}(2021)\citenamefont
  {Kavokine}, \citenamefont {Netz},\ and\ \citenamefont
  {Bocquet}}]{kavokin.n:2021}%
  \BibitemOpen
  \bibfield  {author} {\bibinfo {author} {\bibfnamefont {N.}~\bibnamefont
  {Kavokine}}, \bibinfo {author} {\bibfnamefont {R.~R.}\ \bibnamefont {Netz}},
  \ and\ \bibinfo {author} {\bibfnamefont {L.}~\bibnamefont {Bocquet}},\
  }\href@noop {} {\bibfield  {journal} {\bibinfo  {journal} {Ann. Rev. Fluid
  Mech.}\ }\textbf {\bibinfo {volume} {53}},\ \bibinfo {pages} {377} (\bibinfo
  {year} {2021})}\BibitemShut {NoStop}%
\bibitem [{\citenamefont {Connor}\ and\ \citenamefont
  {Horn}(2001)}]{connor.jn:2001}%
  \BibitemOpen
  \bibfield  {author} {\bibinfo {author} {\bibfnamefont {J.~N.}\ \bibnamefont
  {Connor}}\ and\ \bibinfo {author} {\bibfnamefont {R.~G.}\ \bibnamefont
  {Horn}},\ }\href@noop {} {\bibfield  {journal} {\bibinfo  {journal}
  {Langmuir}\ }\textbf {\bibinfo {volume} {17}},\ \bibinfo {pages} {7194}
  (\bibinfo {year} {2001})}\BibitemShut {NoStop}%
\bibitem [{\citenamefont {Clasohm}\ \emph {et~al.}(2006)\citenamefont
  {Clasohm}, \citenamefont {Chen}, \citenamefont {Knoll}, \citenamefont
  {Vinogradova},\ and\ \citenamefont {Horn}}]{clasohm.ly:2006}%
  \BibitemOpen
  \bibfield  {author} {\bibinfo {author} {\bibfnamefont {L.~Y.}\ \bibnamefont
  {Clasohm}}, \bibinfo {author} {\bibfnamefont {M.}~\bibnamefont {Chen}},
  \bibinfo {author} {\bibfnamefont {W.}~\bibnamefont {Knoll}}, \bibinfo
  {author} {\bibfnamefont {O.~I.}\ \bibnamefont {Vinogradova}}, \ and\ \bibinfo
  {author} {\bibfnamefont {R.~G.}\ \bibnamefont {Horn}},\ }\href@noop {}
  {\bibfield  {journal} {\bibinfo  {journal} {J. Phys. Chem. B}\ }\textbf
  {\bibinfo {volume} {110}},\ \bibinfo {pages} {25931} (\bibinfo {year}
  {2006})}\BibitemShut {NoStop}%
\bibitem [{\citenamefont {Yaroshchuk}\ \emph {et~al.}(2009)\citenamefont
  {Yaroshchuk}, \citenamefont {Boiko},\ and\ \citenamefont
  {Makovetskiy}}]{yaroshchuk.a:2009}%
  \BibitemOpen
  \bibfield  {author} {\bibinfo {author} {\bibfnamefont {A.}~\bibnamefont
  {Yaroshchuk}}, \bibinfo {author} {\bibfnamefont {Y.}~\bibnamefont {Boiko}}, \
  and\ \bibinfo {author} {\bibfnamefont {A.}~\bibnamefont {Makovetskiy}},\
  }\href@noop {} {\bibfield  {journal} {\bibinfo  {journal} {Langmuir}\
  }\textbf {\bibinfo {volume} {25}},\ \bibinfo {pages} {9605} (\bibinfo {year}
  {2009})}\BibitemShut {NoStop}%
\bibitem [{\citenamefont {Derjaguin}\ and\ \citenamefont
  {Landau}(1941)}]{derjaguin.bv:1941}%
  \BibitemOpen
  \bibfield  {author} {\bibinfo {author} {\bibfnamefont {B.~V.}\ \bibnamefont
  {Derjaguin}}\ and\ \bibinfo {author} {\bibfnamefont {L.~D.}\ \bibnamefont
  {Landau}},\ }\href@noop {} {\bibfield  {journal} {\bibinfo  {journal} {Acta
  Physicochimica U.R.S.S.}\ }\textbf {\bibinfo {volume} {14}},\ \bibinfo
  {pages} {633} (\bibinfo {year} {1941})}\BibitemShut {NoStop}%
\bibitem [{\citenamefont {Silkina}\ \emph {et~al.}(2021)\citenamefont
  {Silkina}, \citenamefont {Bag},\ and\ \citenamefont
  {Vinogradova}}]{silkina.ef:2021}%
  \BibitemOpen
  \bibfield  {author} {\bibinfo {author} {\bibfnamefont {E.~F.}\ \bibnamefont
  {Silkina}}, \bibinfo {author} {\bibfnamefont {N.}~\bibnamefont {Bag}}, \ and\
  \bibinfo {author} {\bibfnamefont {O.~I.}\ \bibnamefont {Vinogradova}},\
  }\href@noop {} {\bibfield  {journal} {\bibinfo  {journal} {J. Chem. Phys.}\
  }\textbf {\bibinfo {volume} {154}},\ \bibinfo {pages} {164701} (\bibinfo
  {year} {2021})}\BibitemShut {NoStop}%
\bibitem [{\citenamefont {Markovich}\ \emph {et~al.}(2016)\citenamefont
  {Markovich}, \citenamefont {Andelman},\ and\ \citenamefont
  {Podgornik}}]{markovich.t:2016}%
  \BibitemOpen
  \bibfield  {author} {\bibinfo {author} {\bibfnamefont {T.}~\bibnamefont
  {Markovich}}, \bibinfo {author} {\bibfnamefont {D.}~\bibnamefont {Andelman}},
  \ and\ \bibinfo {author} {\bibfnamefont {R.}~\bibnamefont {Podgornik}},\
  }\href@noop {} {\bibfield  {journal} {\bibinfo  {journal} {EPL}\ }\textbf
  {\bibinfo {volume} {113}},\ \bibinfo {pages} {26004} (\bibinfo {year}
  {2016})}\BibitemShut {NoStop}%
\bibitem [{\citenamefont {Karnik}\ \emph
  {et~al.}(2005{\natexlab{a}})\citenamefont {Karnik}, \citenamefont {Fan},
  \citenamefont {Yue}, \citenamefont {Li}, \citenamefont {Yang},\ and\
  \citenamefont {Majumdar}}]{karnik.r:2005}%
  \BibitemOpen
  \bibfield  {author} {\bibinfo {author} {\bibfnamefont {R.}~\bibnamefont
  {Karnik}}, \bibinfo {author} {\bibfnamefont {R.}~\bibnamefont {Fan}},
  \bibinfo {author} {\bibfnamefont {M.}~\bibnamefont {Yue}}, \bibinfo {author}
  {\bibfnamefont {D.}~\bibnamefont {Li}}, \bibinfo {author} {\bibfnamefont
  {P.}~\bibnamefont {Yang}}, \ and\ \bibinfo {author} {\bibfnamefont
  {A.}~\bibnamefont {Majumdar}},\ }\href@noop {} {\bibfield  {journal}
  {\bibinfo  {journal} {Nano Lett.}\ }\textbf {\bibinfo {volume} {5}},\
  \bibinfo {pages} {943} (\bibinfo {year} {2005}{\natexlab{a}})}\BibitemShut
  {NoStop}%
\bibitem [{\citenamefont {Karnik}\ \emph
  {et~al.}(2005{\natexlab{b}})\citenamefont {Karnik}, \citenamefont
  {Castelino}, \citenamefont {Fan}, \citenamefont {Yang},\ and\ \citenamefont
  {Majumdar}}]{karnik.r:2005b}%
  \BibitemOpen
  \bibfield  {author} {\bibinfo {author} {\bibfnamefont {R.}~\bibnamefont
  {Karnik}}, \bibinfo {author} {\bibfnamefont {K.}~\bibnamefont {Castelino}},
  \bibinfo {author} {\bibfnamefont {R.}~\bibnamefont {Fan}}, \bibinfo {author}
  {\bibfnamefont {P.}~\bibnamefont {Yang}}, \ and\ \bibinfo {author}
  {\bibfnamefont {A.}~\bibnamefont {Majumdar}},\ }\href@noop {} {\bibfield
  {journal} {\bibinfo  {journal} {Nano Lett.}\ }\textbf {\bibinfo {volume}
  {5}},\ \bibinfo {pages} {1638} (\bibinfo {year}
  {2005}{\natexlab{b}})}\BibitemShut {NoStop}%
\bibitem [{\citenamefont {Cervera}(2006)}]{cervera.j:2006}%
  \BibitemOpen
  \bibfield  {author} {\bibinfo {author} {\bibfnamefont {J.}~\bibnamefont
  {Cervera}},\ }\href@noop {} {\bibfield  {journal} {\bibinfo  {journal} {J.
  Chem. Phys.}\ }\textbf {\bibinfo {volume} {124}},\ \bibinfo {pages} {104706}
  (\bibinfo {year} {2006})}\BibitemShut {NoStop}%
\bibitem [{\citenamefont {Dal~Cegnio}\ and\ \citenamefont
  {Pagonabarraga}(2019)}]{dalcengio.s:2019}%
  \BibitemOpen
  \bibfield  {author} {\bibinfo {author} {\bibfnamefont {S.}~\bibnamefont
  {Dal~Cegnio}}\ and\ \bibinfo {author} {\bibfnamefont {I.}~\bibnamefont
  {Pagonabarraga}},\ }\href@noop {} {\bibfield  {journal} {\bibinfo  {journal}
  {J. Chem. Phys.}\ }\textbf {\bibinfo {volume} {151}},\ \bibinfo {pages}
  {044707} (\bibinfo {year} {2019})}\BibitemShut {NoStop}%
\bibitem [{\citenamefont {Ren}\ and\ \citenamefont
  {Stein}(2008)}]{ren.yq:2008}%
  \BibitemOpen
  \bibfield  {author} {\bibinfo {author} {\bibfnamefont {Y.~Q.}\ \bibnamefont
  {Ren}}\ and\ \bibinfo {author} {\bibfnamefont {D.}~\bibnamefont {Stein}},\
  }\href@noop {} {\bibfield  {journal} {\bibinfo  {journal} {Nanotechnology}\
  }\textbf {\bibinfo {volume} {19}},\ \bibinfo {pages} {195707} (\bibinfo
  {year} {2008})}\BibitemShut {NoStop}%
\bibitem [{\citenamefont {Zhang}\ \emph {et~al.}(2021)\citenamefont {Zhang},
  \citenamefont {Wen},\ and\ \citenamefont {Jiang}}]{zhang.z:2021}%
  \BibitemOpen
  \bibfield  {author} {\bibinfo {author} {\bibfnamefont {Z.}~\bibnamefont
  {Zhang}}, \bibinfo {author} {\bibfnamefont {L.}~\bibnamefont {Wen}}, \ and\
  \bibinfo {author} {\bibfnamefont {L.}~\bibnamefont {Jiang}},\ }\href@noop {}
  {\bibfield  {journal} {\bibinfo  {journal} {Nat. Rev. Mater}\ ,\ \bibinfo
  {pages} {https://doi.org/10.1038/s41578}} (\bibinfo {year}
  {2021})}\BibitemShut {NoStop}%
\bibitem [{\citenamefont {Wang}\ \emph {et~al.}(2021)\citenamefont {Wang},
  \citenamefont {Wang}, \citenamefont {Patel}, \citenamefont {Lin},\ and\
  \citenamefont {Elimelech}}]{wang.l:2021}%
  \BibitemOpen
  \bibfield  {author} {\bibinfo {author} {\bibfnamefont {L.}~\bibnamefont
  {Wang}}, \bibinfo {author} {\bibfnamefont {Z.}~\bibnamefont {Wang}}, \bibinfo
  {author} {\bibfnamefont {S.~K.}\ \bibnamefont {Patel}}, \bibinfo {author}
  {\bibfnamefont {S.}~\bibnamefont {Lin}}, \ and\ \bibinfo {author}
  {\bibfnamefont {M.}~\bibnamefont {Elimelech}},\ }\href@noop {} {\bibfield
  {journal} {\bibinfo  {journal} {ACS Nano}\ }\textbf {\bibinfo {volume}
  {15}},\ \bibinfo {pages} {4093} (\bibinfo {year} {2021})}\BibitemShut
  {NoStop}%
\end{thebibliography}%

\end{document}